\documentclass[iop,apj]{emulateapj}
\usepackage{amsmath }
\usepackage{epstopdf}

\shortauthors{Ruan et al.}
\shorttitle{Archival Search for Changing-Look Quasars} 
\begin{document}
\title{Towards an Understanding of Changing-Look Quasars: \\An Archival Spectroscopic Search in SDSS}
\author{John~J.~Ruan\altaffilmark{1,2},
 Scott~F.~Anderson\altaffilmark{2}, 
 Sabrina~L.~Cales\altaffilmark{3,4},
 Michael~Eracleous\altaffilmark{5},
 Paul~J.~Green\altaffilmark{6}, 
 Eric~Morganson\altaffilmark{6}, 
 Jessie~C.~Runnoe\altaffilmark{5},
 Yue~Shen\altaffilmark{7,8},
 Tessa~D.~Wilkinson\altaffilmark{2},
 Michael~R.~Blanton\altaffilmark{9},
 Tom~Dwelly\altaffilmark{10},  
 Antonis~Georgakakis\altaffilmark{10},  
 Jenny~E.~Greene\altaffilmark{11}, 
 Stephanie~M.~LaMassa\altaffilmark{3},
 Andrea~Merloni\altaffilmark{10}, 
 Donald~P.~Schneider\altaffilmark{5}
}
\altaffiltext{1}{ Corresponding author: jruan@astro.washington.edu} 
\altaffiltext{2}{Department of Astronomy, University of
Washington, Box 351580, Seattle, WA 98195, USA}
\altaffiltext{3}{Yale Center for Astronomy and Astrophysics, Physics Department, 
Yale University, New Haven, CT, 06511}
\altaffiltext{4}{Department of Astronomy, University of Concepcion, Concepcion, Chile}
\altaffiltext{5}{Department of Astronomy \& Astrophysics and Institute for Gravitation and the Cosmos, 
525 Davey Lab, The Pennsylvania State University, University Park, PA 16802, USA}
\altaffiltext{6}{Harvard Smithsonian Center for Astrophysics, 
60 Garden St, Cambridge, MA 02138, USA}
\altaffiltext{7}{Kavli Institute for Astronomy and Astrophysics, Peking University,
Beijing 100871, China}
\altaffiltext{8}{Carnegie Observatories, 813 Santa Barbara Street, Pasadena, CA
91101, USA}
\altaffiltext{9}{Center for Cosmology and Particle Physics,
New York University, Department of Physics, New York University,
4 Washington Pl, New York, NY 10003}
\altaffiltext{10}{Max-Planck-Institut fur extraterrestrische Physik (MPE), 
Giessenbachstrasse 1, D-85748, Garching bei M\"unchen, Germany}
\altaffiltext{11}{Department of Astrophysical Sciences, Princeton University, Peyton Hall, 
Princeton, NJ 08544, USA}

\begin{abstract}
	The uncertain origin of the recently-discovered `changing-looking' quasar phenomenon -- in which a luminous quasar dims significantly to a quiescent state in repeat spectroscopy over $\sim$10 year timescales -- may present unexpected challenges to our understanding of quasar accretion. To better understand this phenomenon, we take a first step to building a sample of changing-look quasars with a systematic but simple archival search for these objects in the Sloan Digital Sky Survey Data Release 12. By leveraging the $>$10 year baselines for objects with repeat spectroscopy, we uncover two new changing-look quasars, and a third discovered previously. Decomposition of the multi-epoch spectra and analysis of the broad emission lines suggest that the quasar accretion disk emission dims due to rapidly decreasing accretion rates (by factors of $\gtrsim$2.5), while disfavoring changes in intrinsic dust extinction for the two objects where these analyses are possible. Broad emission line energetics also support intrinsic dimming of quasar emission as the origin for this phenomenon rather than transient tidal disruption events or supernovae. Although our search criteria included quasars at all redshifts and transitions from either quasar-like to galaxy-like states or the reverse, all of the clear cases of changing-look quasars discovered were at relatively low-redshift ($z \sim 0.2 - 0.3$) and only exhibit quasar-like to galaxy-like transitions.
\end{abstract}

\section{Introduction}
	The quasar phenomenon is thought to be a relatively brief stage of galaxy evolution involving rapid accretion onto the central supermassive black hole \citep[SMBH; e.g.,][]{salpeter64,lyndenbell69, rees78}. Observational constraints on lifetimes show that quasar phases in galaxies generally last for a total of $10^{7-8}$ years \citep{martini01, kelly10}, after which the accretion rate drops dramatically and the active nucleus transitions to a low-luminosity active galactic nucleus (AGN) or quiescent galaxy state \citep{churazov05}. Cosmological simulations of galaxy formation that include sub-grid models of SMBH growth and feedback have suggested that the accretion history of SMBHs may be episodic, where luminous quasar phases are regulated by quasar feedback processes \citep[e.g.,][]{dimatteo05,hopkins05,springel05}. Although the exact characteristics of quasar light curves over cosmic time are difficult to infer observationally, indirect arguments based on AGN populations \citep{schawinski15} and properties of their host-galaxies \citep{hickox14} have also suggested that AGNs dramatically `flicker' in luminosity between luminous quasar and quiescent galaxy phases. However, direct observations of such transitions in luminous quasars have thus far been scarce. 

	It has been suggested that the transition from quasars to low-luminosity AGN or quiescent galaxies may not be observable in individual objects due to the long timescales expected for this process. Such dramatic changes in the accretion state are commonly observed in X-ray binaries, which can undergo spectral state transitions between the high-luminosity/soft-spectrum and low-luminosity/hard-spectrum states in the X-rays \citep[e.g.,][]{homan05}. Scaling the hours-long timescales observed for these spectral state transitions in X-ray binaries to $\sim$10$^8$ $M_\odot$ SMBHs predicts transition timescales in quasars of $\sim$10$^{4-5}$ years \citep{sobolewska11}. Indirect evidence for a luminous quasar transition in an individual object was previously provided by observations of Hanny's Voorwerp, a serendipitously discovered ionized emission-line gas cloud lying 
$\sim$20 kpc away from the quiescent galaxy IC2497 \citep{lintott09}. Based on multi-wavelength observations, it is argued that this gas cloud could only have been ionized by an AGN continuum with luminosity $\gtrsim$10$^{45}$ erg s$^{-1}$; this implies that the nearby quiescent galaxy was recently in a quasar state, with a transition timescale of $\sim$10$^{4}$ years and consistent with expectations \citep{schawinski10, keel12a}. Since this discovery, many other candidate fading AGNs with extended emission-line regions have been found and investigated, resulting in similar inferred transition timescales \citep[e.g.,][]{keel12b, keel15}.

	Previously, transitions of Seyfert 1 galaxies to Seyfert 1.8/1.9 (and/or vice versa) have been directly observed in repeat optical spectroscopy over timescales of $\sim$10 years \citep{goodrich95, shappee14, denney14}. \citet{goodrich95} showed that the origin of the observed transition in some of these AGNs are consistent with intrinsic changes in the AGN continuum emission, while variations in dust obscuration along the line of sight is favored for others, although these two effects may occur in concert if dust is embedded in the narrow-line region gas \citep{netzer93}. Similar behavior has also been previously observed in the X-ray spectra of Seyfert galaxies (termed ``changing-look AGN"), also interpreted as being due to either dramatic changes in obscuration or intrinsic changes in the nuclear emission \citep[e.g.,][]{guainazzi02, matt03, puccetti2007, risaliti09, marchese12}. The recent discovery of the first changing-look (CL) quasar by \citet{lamassa15} extends this transitional phenomenon to AGNs in new luminosity and redshift regimes (see Figure 1 of \citealt{lamassa15}). Repeat optical spectroscopy of this luminous quasar (SDSS J015957.64+003310.5, hereafter J0159+0033) shows a dramatic decrease in the quasar continuum emission, accompanied by disappearance of the broad H$\beta$ line and strong dimming of the broad H$\alpha$ line, showing that a simple orientation-based view of AGN unification is incomplete. Surprisingly, the observed transition in this CL quasar occurred over rest-frame timescales of $\sim$7 years; this is much shorter than the $\sim$10$^4$ year timescales expected for this transition to occur from previous arguments based on X-ray binaries and extended emission line regions surrounding quiescent galaxies.
	
	The origin of changing-look behavior in luminous quasars is uncertain. \citet{lamassa15} demonstrated that the observed dimming of the quasar continuum in J0159+0033 coincides with broadening of the broad Balmer emission lines, such that the derived black hole mass (estimated through single-epoch spectroscopic black hole mass methods) is preserved. This behavior is consistent with intrinsic dimming of the quasar continuum emission, while a scenario in which the continuum and emission line dimming is caused by an increase in dust extinction is disfavored through modeling of the spectral changes. Furthermore, \citet{lamassa15} also argue that obscuration by a dust cloud outside the broad line region in a circular Keplerian orbit is unlikely since its crossing time across the broad line region would be much longer than the observed transition timescale.
	
	If the dimming of quasar emission in CL quasars is intrinsic, then the observed behavior can be caused by dramatic changes in the accretion flow, which may occur during transitions between radiatively efficient and inefficient accretion regimes \citep{ichimaru77,rees82,narayan94}. Thermal and dynamical instabilities in the accretion disk may also produce strong changes in the disk emission on even shorter timescales \citep{lin86, siemiginowska96}. \citet{merloni15} argued that the \citet{lamassa15} CL quasar may instead be a transient stellar tidal disruption event (TDE) near the central SMBH \citep[see also][]{eracleous95}, which would cause a luminous nuclear flare, followed by a slow dimming over the few-years timescales observed. This scenario is supported by their image-differencing light curves of this CL quasar, which show that the time evolution of the broadband nuclear emission is consistent with that expected from TDEs. In any of these scenarios, CL quasars represent an intriguing new phenomenon that can provide unique insights to AGN accretion and structure, and warrant additional investigation.
	
	Since the discovery of \citet{lamassa15}, at least two more spectroscopic CL quasars have been serendipitously discovered \citep{runnoe16}, primarily through visual inspection of early spectroscopic data from the Time-Domain Spectroscopic Survey \citep{morganson15} in the Sloan Digital Sky Survey IV (SDSS IV). These results suggest that CL quasars may be surprisingly common, and can be found by mining spectral data sets with large numbers of repeat quasar and galaxy spectra over a sufficiently long baseline. Motivated by these results, we present a simple archival search for CL quasars in SDSS-I/II/III Data Release 12 \citep[DR12,][]{alam15}, which includes a total of $\sim$$4\times10^6$ optical/near-IR spectra over $\sim$10,000 deg$^2$ of sky. This data set includes a significant number of repeat spectra spanning a $\gtrsim$10 year baseline, and many of its various science programs specifically targeted quasars and galaxies. We aim to cast a wide net in this investigation, and include quasar and galaxies at all redshifts while remaining impartial in our search with regard to whether each object transitions from a quasar-like to galaxy-like state or vice versa; this approach could not only yield a sample of CL quasars, but also potentially a spectroscopic transition from quiescent galaxy to luminous quasar, which would have intriguing implications for their origin.

	The outline of this paper is as follows: Section 2 presents the data sets and criteria used in our search. In Section 3, we describe the changing look quasars found in our search and modeling of their broad emission lines. Section 4 evaluates evidence from our sample favoring various scenarios for the origin of CL quasars. We summarize and conclude in Section 5. Throughout this paper, we assume a standard $\Lambda$CDM cosmology with $\Omega_\mathrm{m} = 0.309$, $\Omega_\Lambda = 0.691$, and $H_0 = 67.7$ km s$^{-1}$ Mpc$^{-1}$, consistent with the \emph{Planck} full-mission results of \citet{planck15}.
	
\begin{deluxetable*}{cccccccccc}
\tablecolumns{12}
\tablewidth{0pt}
\tablecaption{Measured SDSS spectral properties of the changing-look quasars in our sample.}
\tablehead {\colhead{SDSS}  & \colhead{$z$} &\colhead{MJD} & \colhead{H$\alpha$ FWHM$^\textrm{a}$} & \colhead{log$_{10}$$L_\mathrm{H\alpha}$$^\textrm{a}$} & \colhead{H$\beta$ FWHM$^\textrm{a}$} & \colhead{log$_{10}$$L_\mathrm{H\beta}$$^\textrm{a}$} & \colhead{log$_{10}$$\lambda L_\mathrm{5100}$}\\ 
\colhead{Object} & & & \colhead{[km s$^{-1}$]} & \colhead{[erg s$^{-1}$]} & \colhead{[km s$^{-1}$]} & \colhead{[erg s$^{-1}$]} & \colhead{[erg s$^{-1}$]}
}
\startdata
\phantom{0}J015957.64$+$003310.5$^\textrm{b}$ & 0.312 & 51871 & 3788 $\pm$ 163 & 42.36 $\pm$ 0.04 & 4714 $\pm$ 682 & 41.88 $\pm$ 0.63 & 43.52 $\pm$ 0.05 \\
& & 55201 & 5954 $\pm$ 857 & 41.72 $\pm$ 0.11 & .... & $<$41.32$^\textrm{c}$  & 43.27 $\pm$ 0.06 \\
\rule{0pt}{2.5ex}  


J012648.08$-$083948.0 &  0.198 & 52163 & 4121 $\pm$ 223 & 42.00 $\pm$ 0.04 & 4297 $\pm$ 1165 & 41.55 $\pm$ 0.20 & 43.43 $\pm$ 0.03 \\
& & 54465 & .... & $<$40.20$^\textrm{d}$, $<$40.51$^\textrm{e}$ & ... & $<$40.17$^\textrm{f}$, $<$40.28$^\textrm{g}$ & $<$42.30$^\textrm{c}$ \\
\rule{0pt}{2.5ex}  

J233602.98$+$001728.7 & 0.243 & \phantom{0}52096$^\textrm{h}$ & 6289 $\pm$ 1180 & 41.86 $\pm$ 0.20 & 6993 $\pm$ 2271 & 41.28 $\pm$ 0.20 & 43.04 $\pm$ 0.09 \\
& & 55449 & 7209 $\pm$ 1367 & 41.48 $\pm$ 0.22 & ... & $<$40.60$^\textrm{c}$ & 42.56 $\pm$ 0.18  \\
\enddata
\tablenotetext{}{}
\tablenotetext{a}{These measurements of the luminosities and widths are for the broad components of these Balmer lines.} 
\tablenotetext{b}{Changing-look quasar previously found by \citet{lamassa15} and also disused in \citet{merloni15}.} 
\tablenotetext{c}{5$\sigma$ upper limit assuming FWHM predicted from H$\alpha$, see Section 3.2.} 
\tablenotetext{d}{5$\sigma$ upper limit assuming FWMH of 7960 km s$^{-1}$, see Section 3.2.} 
\tablenotetext{e}{5$\sigma$ upper limit assuming FWMH of 4121 km s$^{-1}$, see Section 3.2.} 
\tablenotetext{f}{5$\sigma$ upper limit assuming FWMH of 8920 km s$^{-1}$, see Section 3.2.} 
\tablenotetext{g}{5$\sigma$ upper limit assuming FWMH of 4297 km s$^{-1}$, see Section 3.2.} 
\tablenotetext{h}{This MJD is the mean of four closely-spaced epochs of spectra that have been stacked (see discussion in Section 3.1).} 
\end{deluxetable*}

\begin{deluxetable*}{cccccc}
\tablecolumns{12}
\tablewidth{0pt}
\tablecaption{Inferred SDSS spectral properties of the changing-look quasars in our sample.}
\tablehead {\colhead{SDSS} &\colhead{MJD} & \colhead{log$_{10}$$M_\mathrm{BH, H\alpha}$} & \colhead{log$_{10}$$M_\mathrm{BH, H\beta}$} & \colhead{log$_{10}$($L_{\rm{bol}}/L_{\rm{Edd, H\alpha}}$)}  & \colhead{log$_{10}$($L_{\rm{bol}}/L_{\rm{Edd, H\beta}}$)}  \\
\colhead{Object} & & \colhead{[$M_\odot$]} & \colhead{[$M_\odot$]}  & & 
}
\startdata
\phantom{0}J015957.64$+$003310.5$^\textrm{a}$ & 51871 & 7.93 $\pm$ 0.10 & 8.02 $\pm$ 0.33 & $-1.6 \pm 0.1$ & $-1.7 \pm 1.3$ \\
& 55201 & 8.20 $\pm$ 0.26 & .... & $-2.1 \pm 1.6$ & .... \\
\rule{0pt}{2.5ex}  


J012648.08$-$083948.0 & 52163 & 7.96 $\pm$ 0.10 & 7.89 $\pm$ 0.84 & $-1.7 \pm 0.4$ & $-1.7 \pm 0.7$ \\
& 54465 & .... & .... & .... & .... \\
\rule{0pt}{2.5ex}  

J233602.98$+$001728.7 & \phantom{0}52096$^\textrm{b}$ & 8.13 $\pm$ 0.29 & 8.11 $\pm$ 0.53 & $-2.3 \pm 0.4$ & $-2.3 \pm 0.7$ \\
& 55449 & 8.00 $\pm$ 0.30 & ... & $-2.7 \pm 2.6$ & ...  \\
\enddata
\tablenotetext{}{}
\tablenotetext{a}{Changing-look quasar previously found by \citet{lamassa15} and discussed in \citet{merloni15}.} 
\tablenotetext{b}{This MJD is the mean of four closely-spaced epochs of spectra that have been stacked (see discussion in Section 3.1).} 
\end{deluxetable*}

\section{An Archival Spectroscopic Search}
\subsection{Search Criteria}
	We utilize the list of all 4,355,202 spectra in SDSS \citep{york00} DR12, and perform the selection cuts detailed below to produce a final sample of 117 CL quasar candidates. These spectra were taken by the SDSS 2.5m telescope (Gunn et al. 2006) using the SDSS-I/II and Baryon Oscillation Spectroscopic Survey \citep[BOSS,][]{eisenstein11, dawson13} spectrographs \citep{smee13}, and compiled in the `spAll' files produced by the SDSS spectroscopic reduction pipeline \citep{bolton12}. For a CL quasar to show a convincing transition, its multi-epoch spectra must clearly possess quasar-like spectral features in one epoch (power-law continuum and broad emission lines), and galaxy-like features in another epoch (absorption spectra and narrow emission lines if star-formation or nuclear activity is present). While a sophisticated method of detecting this transition in repeat spectra of each object is likely to be more sensitive to subtle changes, our current goal is to search only for the most obvious and convincing cases of CL quasars. Thus, our simple approach relies on the automated SDSS pipeline to classify each spectrum as quasar-like or galaxy-like. Specifically, using the {\tt CLASS} spectral classification provided for each spectrum in our sample (which is based on fitting to a set of galaxy, quasar, and stellar eigenspectra, see \citealt{bolton12}), we create two subsamples: a galaxy-like sample of 2,510,060 spectra where {\tt CLASS = `GALAXY'}, and a quasar-like sample of 587,306 spectra where {\tt CLASS = `QSO'}. In both these subsamples, sky fibers have been removed using the {\tt sourcetype} targeting keyword. Although it is well-known that these automated pipeline classifications occasionally fail to accurately classify the observed spectra, we emphasize that we are focusing on the most obvious and convincing CL quasars, for which the pipeline classifications will suffice for this initial archival search.
	 
	For each spectrum in our galaxy-like sample, we positionally match to the quasar-like sample using a 1$\arcsec$ matching radius to identify objects with repeat spectra and disparate classifications. We impose additional quality-control conditions on this search: the difference in the pipeline redshifts between the two epochs must be $|\Delta$$z| < 0.01$, and the absolute value of the rest-frame time lag between the two epochs is $|\Delta$$t_\mathrm{rest-frame}| > 4$ years. These conditions remove a significant number of false-positives in which the pipeline catastrophically fails to fit the spectrum in one of the epochs. This search results in 180 pairs of repeat spectra of 117 unique objects (a few objects have more than two epochs of spectra), and we visually inspect all spectra of each of these CL quasar candidates.
	
	From the visual inspection, we find three clear cases of CL quasars (listed in Table 1), which include the \citet{lamassa15} CL quasar (J01595+0033) and two additional new convincing cases (J01264$-$0839 and J2336+00172). All of these CL quasars exhibit quasar-like to galaxy-like transitions, and are at relatively low redshifts of $z\sim0.2-0.3$ in comparison to the parent sample of SDSS quasars, which are overwhelmingly at $z>1$ \citep[e.g. see Figure 2 in][]{paris14}. In all three of the CL quasars discovered in our search, the broad H$\beta$ emission disappears while the broad H$\alpha$ emission dims significantly (and disappears in J01264$-$0839), accompanied by dimming of blue quasar UV/optical continuum emission. The median seeing during the exposures of these three CL quasar candidates ranged from 1.43\arcsec to 1.93\arcsec, consistent with typical values for SDSS spectra (i.e. these exposures are not strong outliers). Small differences in the seeing between exposures at this level are not responsible for the observed spectral variability (including broadening of the broad emission lines, see Section 3.2). We additionally find one ambiguous CL quasar where the latest epoch of SDSS spectra appears to show a galaxy-like spectrum at blue wavelengths, but is corrupted at the redder wavelengths. In Appendix A, we present additional recent non-SDSS spectroscopy which demonstrates that this object does not transition to a galaxy-like state at the epoch of the latest spectrum; this behavior instead likely stems from known instrumental issues affecting the particular fiber of the corrupted SDSS spectrum.
	
	 In the visual inspection of the multi-epoch spectra, the vast majority of the false-positives from our search were cases where the pipeline switched between {\tt CLASS = `GALAXY'} and {\tt CLASS = `QSO'} classifications in repeat spectra despite little change in the spectral properties. Often, this occurs for AGN at redshifts of $z \simeq$ 0.4 in which the broad H$\alpha$ emission line is redward of the smaller wavelength coverage of the SDSS spectrograph in the earlier epoch (leading to a {\tt CLASS = `GALAXY'} classification), but visible in the later epoch from the BOSS spectrograph due to its slightly larger wavelength coverage (leading to a {\tt CLASS = `QSO'} classification). As part of the visual inspection, the fiber plugging positions and targeting flags for each pair of repeat spectra were compared to ensure that there is no offset in the fiber position between the two spectra, artificially leading to more host-galaxy emission in the SDSS spectrum. This offset can occur since some SDSS fibers were part of a SDSS program to test the redshift recovery of the spectroscopic pipeline in SDSS-III relative to that in SDSS-I/II, and are identified using the {\tt PROGRAM = `APBIAS'} target flag in the spectra as well as their disparate fiber plugging positions in repeat spectra. One of these objects was recovered in our search, and was removed from our sample. Aside from this offset object and the three CL quasars, all the remaining 113 candidates were rejected because our visual inspection of their repeat spectra did not reveal dramatic disappearance or appearance of broad emission lines. In Appendix B, we further discuss and show examples of the CL quasar candidates rejected in our visual inspection.
	 
\begin{figure*}[t]
\begin{center}
\includegraphics[width=0.7\textwidth]{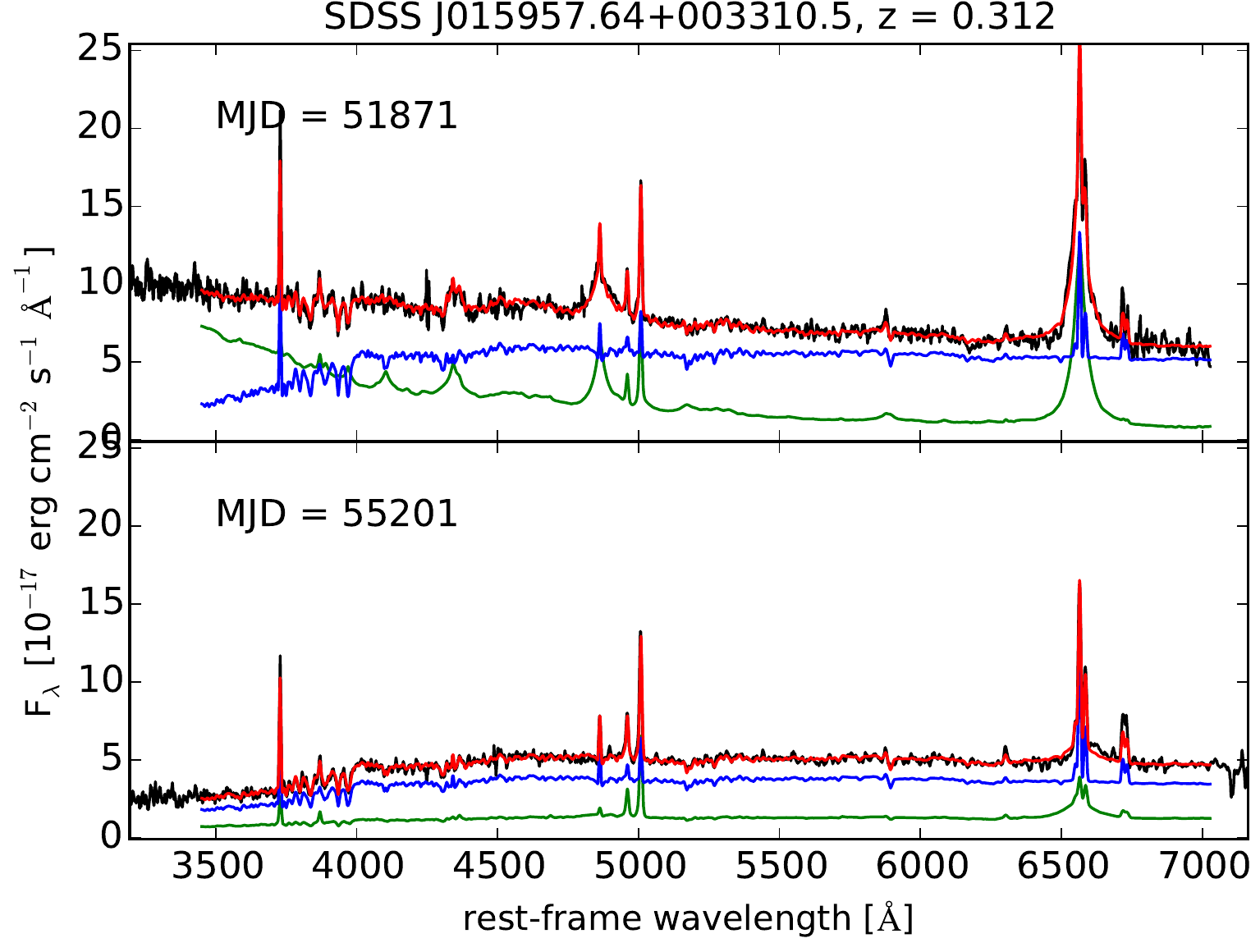} 
\includegraphics[width=0.45\textwidth]{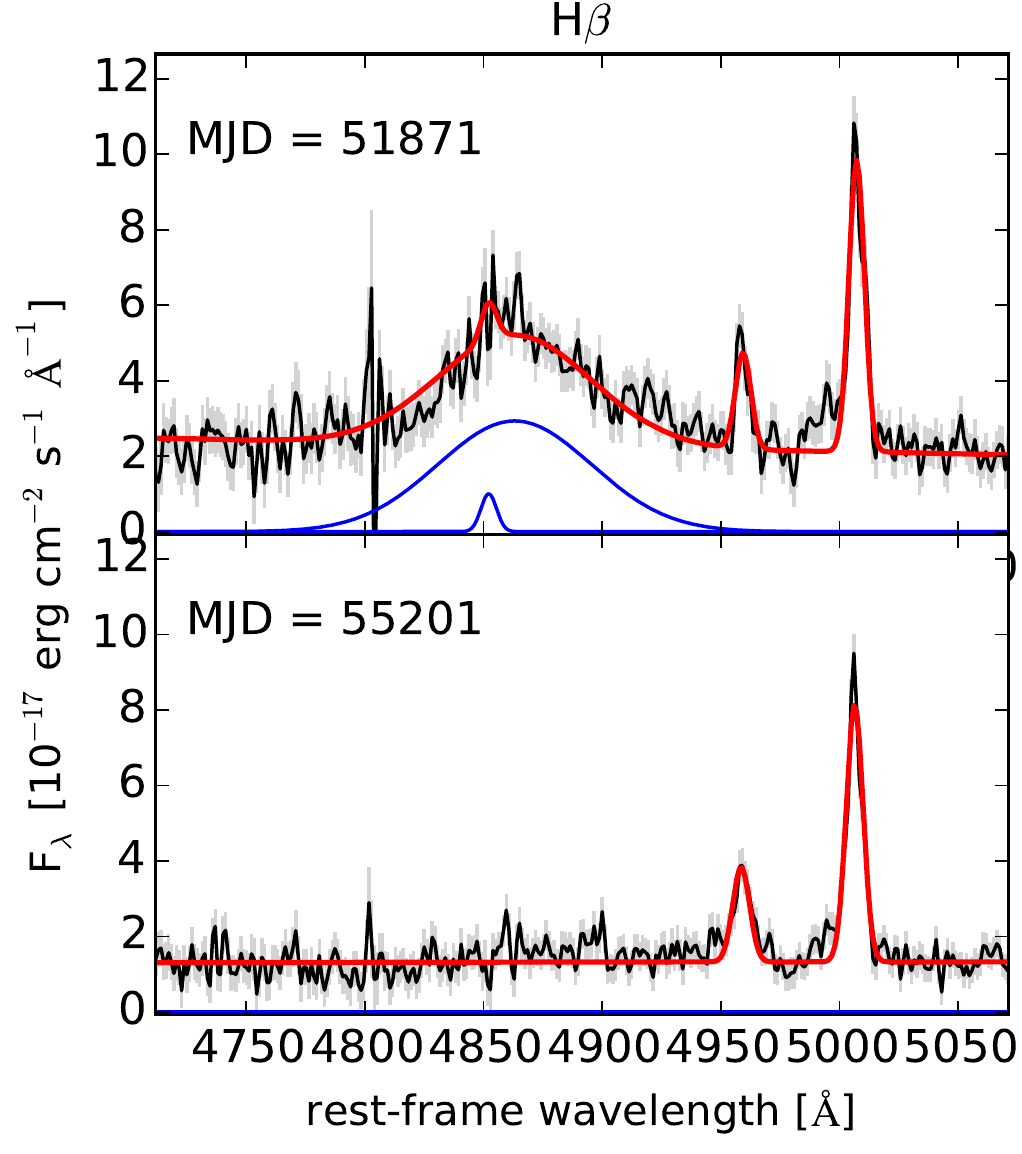} 
\includegraphics[width=0.45\textwidth]{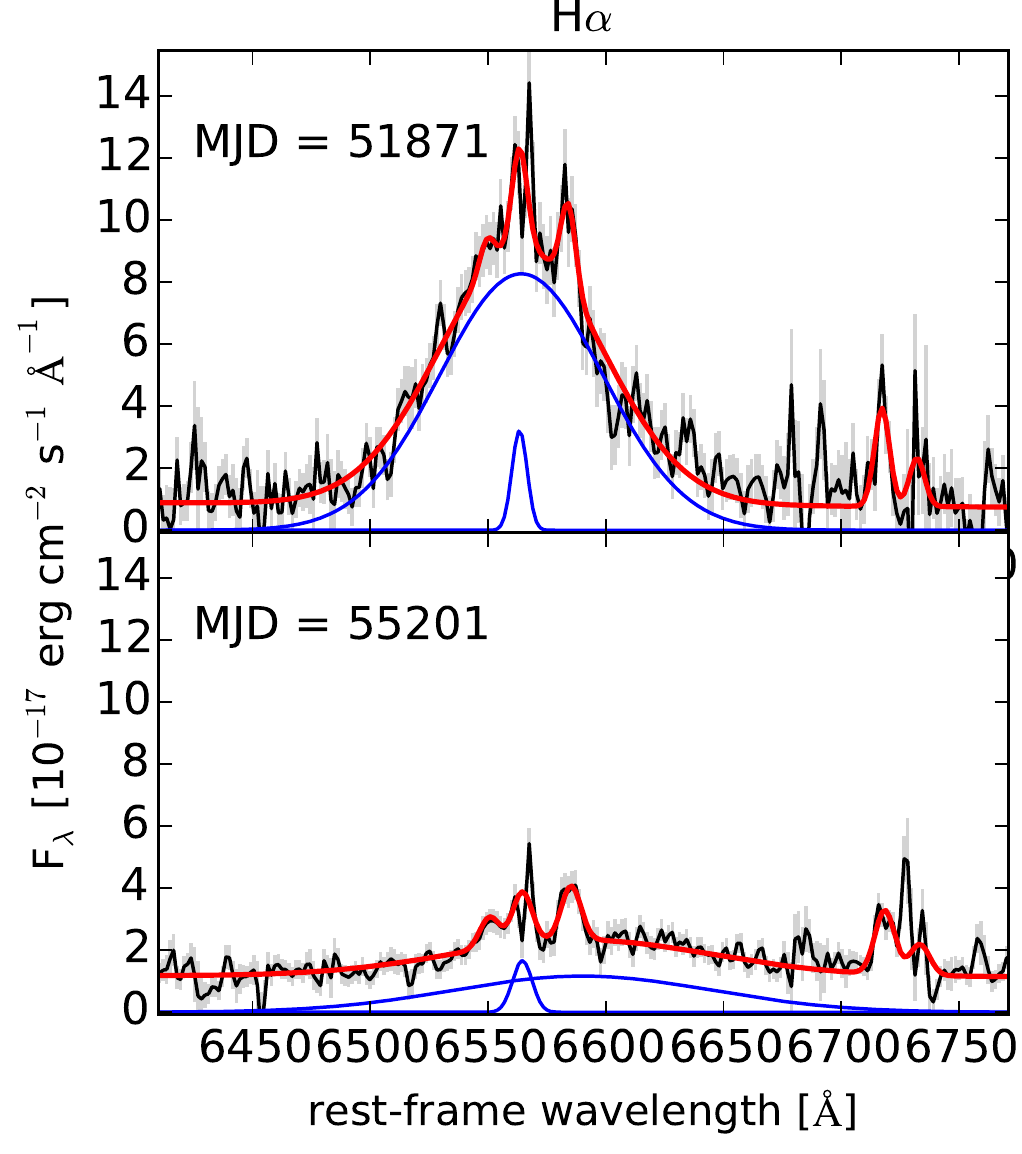} 
\caption{Top: Spectral decomposition for the two epochs of spectra of SDSS J015957.64+003310.5 (see \citealt{lamassa15}), in the rest-frame. The black lines are the observed spectra, and the green and blue lines are the reconstructed quasar and host-galaxy spectra from the eigenspectra decomposition, respectively. The best-fit model to the observed spectrum from the decomposition (i.e. sum of the green and blue lines) is the red line. The dramatic dimming in the quasar continuum and broad Balmer emission lines are consistent with intrinsic dimming of the accretion disk emission rather than dust extinction. Bottom left: Fitting of the H$\beta$ line region in the decomposed quasar spectrum, for the two epochs of spectra. The decomposed quasar spectra are the black lines, the best-fit broad and narrow H$\beta$ emission lines are the blue lines, and the total fits to the decomposed quasar spectra (including quasar continuum and all emission lines) are shown in red. Although narrow emission lines are included in the fit, their amplitudes in the decomposed spectrum are not equivalent to the narrow emission lines in the observed spectrum since they are partially subtracted as part of the host-galaxy spectrum (see Section 3.2). Bottom right: Similar to the bottom left panels, but for the H$\alpha$ emission lines.}
\end{center}
\end{figure*}
\begin{figure*}[t]
\begin{center}
\includegraphics[width=0.7\textwidth]{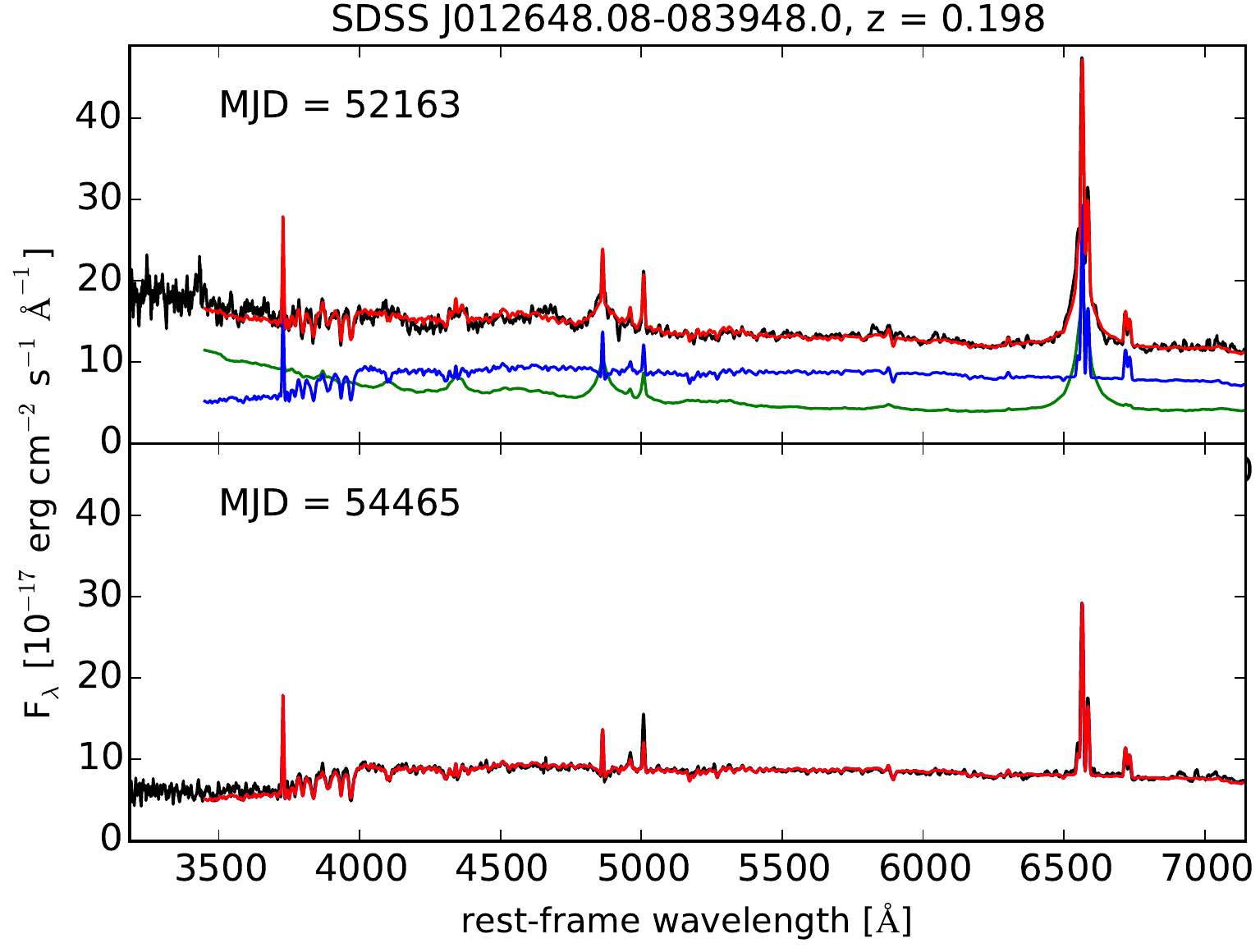} 
\includegraphics[width=0.45\textwidth]{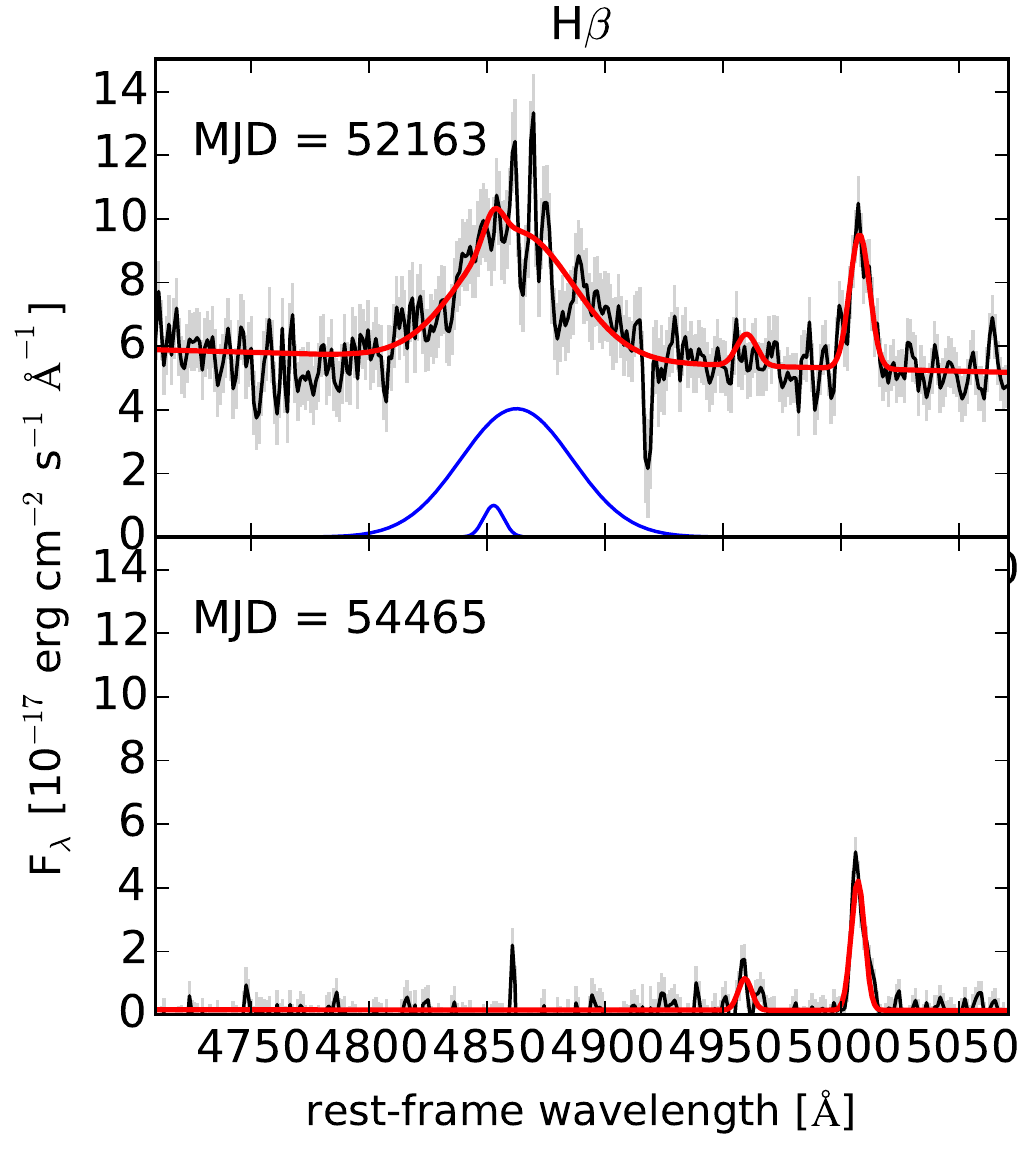} 
\includegraphics[width=0.45\textwidth]{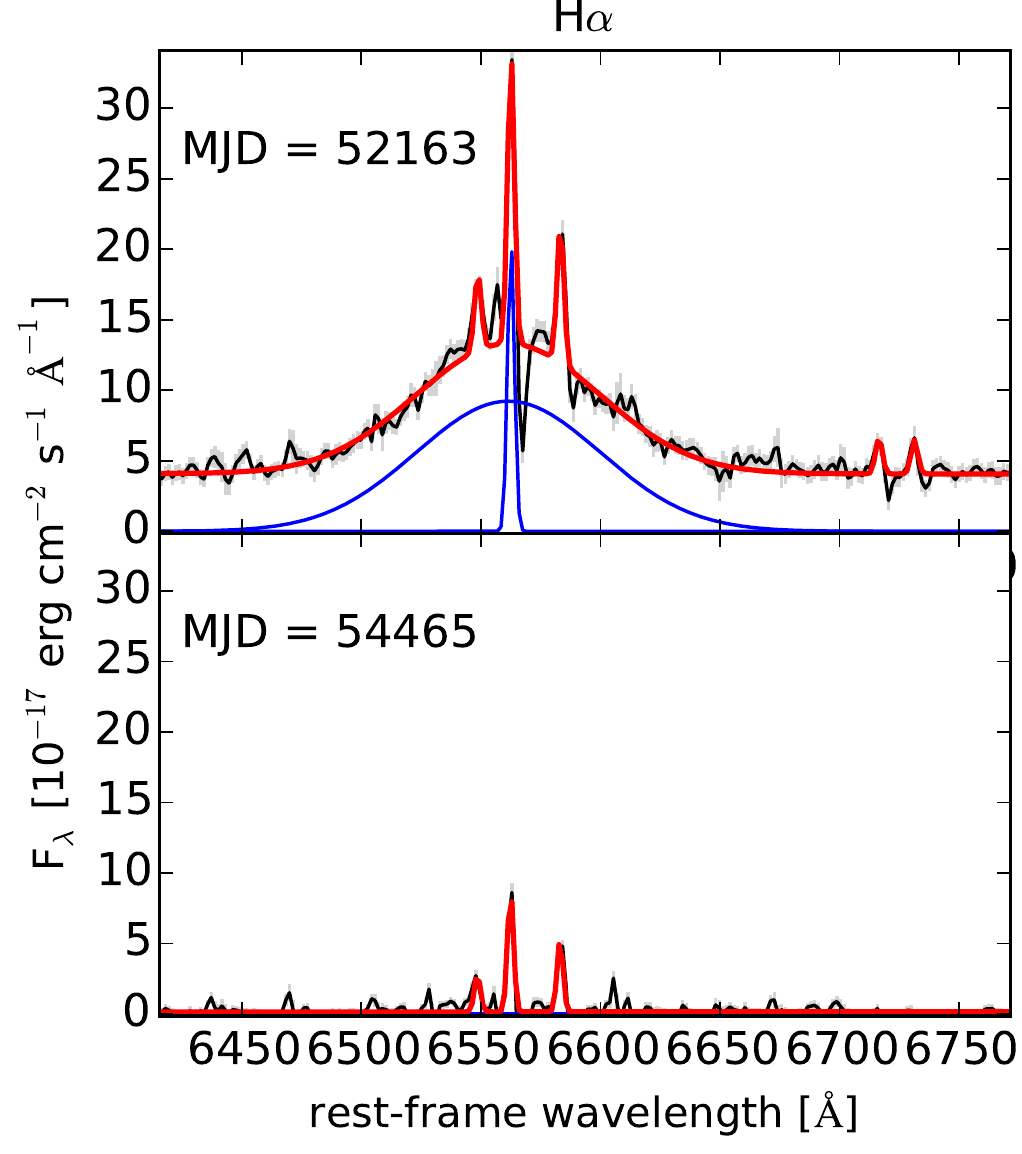} 
\caption{Spectral decomposition for the two epochs of spectra of SDSS J012648.08-083948.0 in the rest-frame, same format as Figure 1.}
\end{center}
\end{figure*}
\begin{figure*}[t]
\begin{center}
\includegraphics[width=0.7\textwidth]{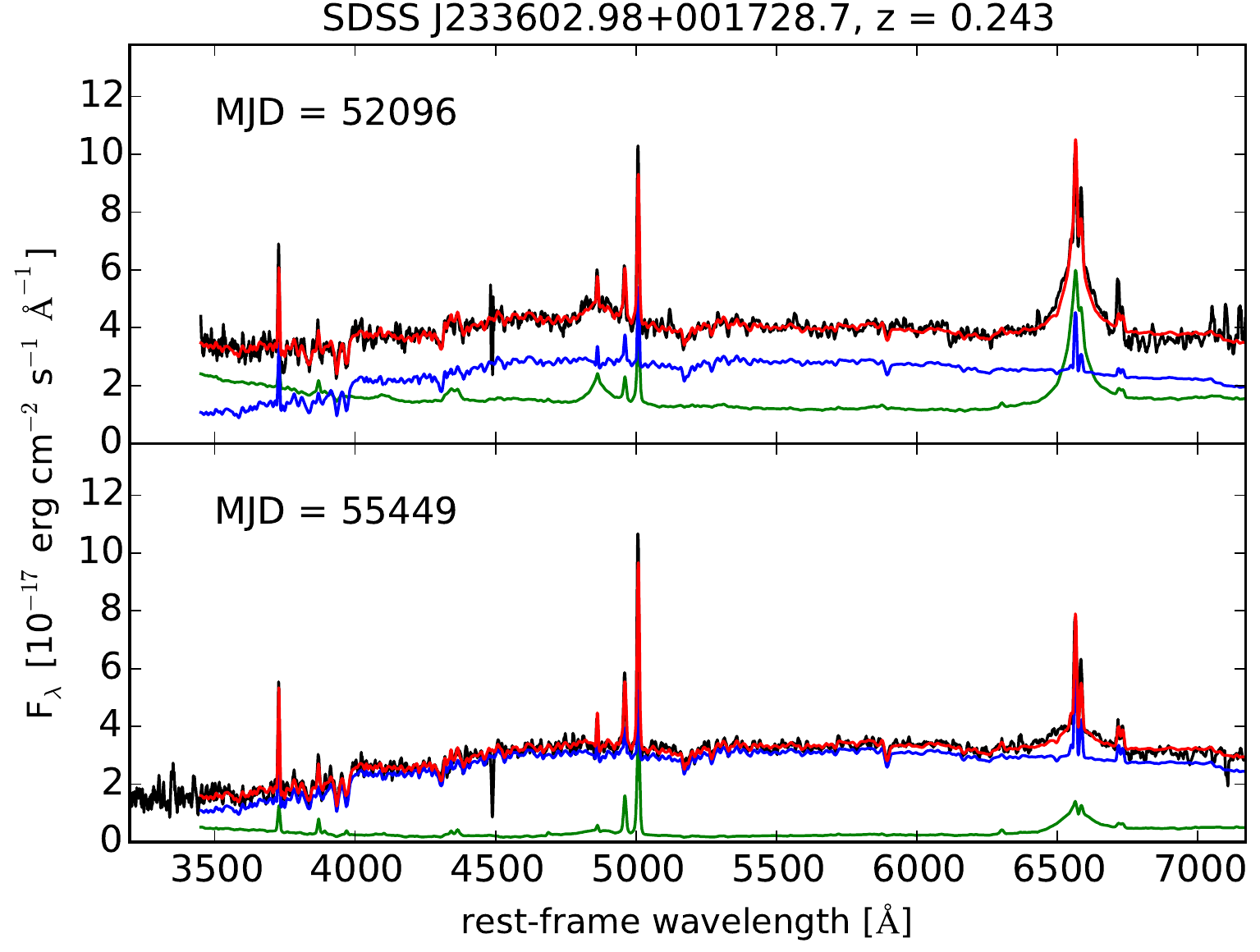} 
\includegraphics[width=0.45\textwidth]{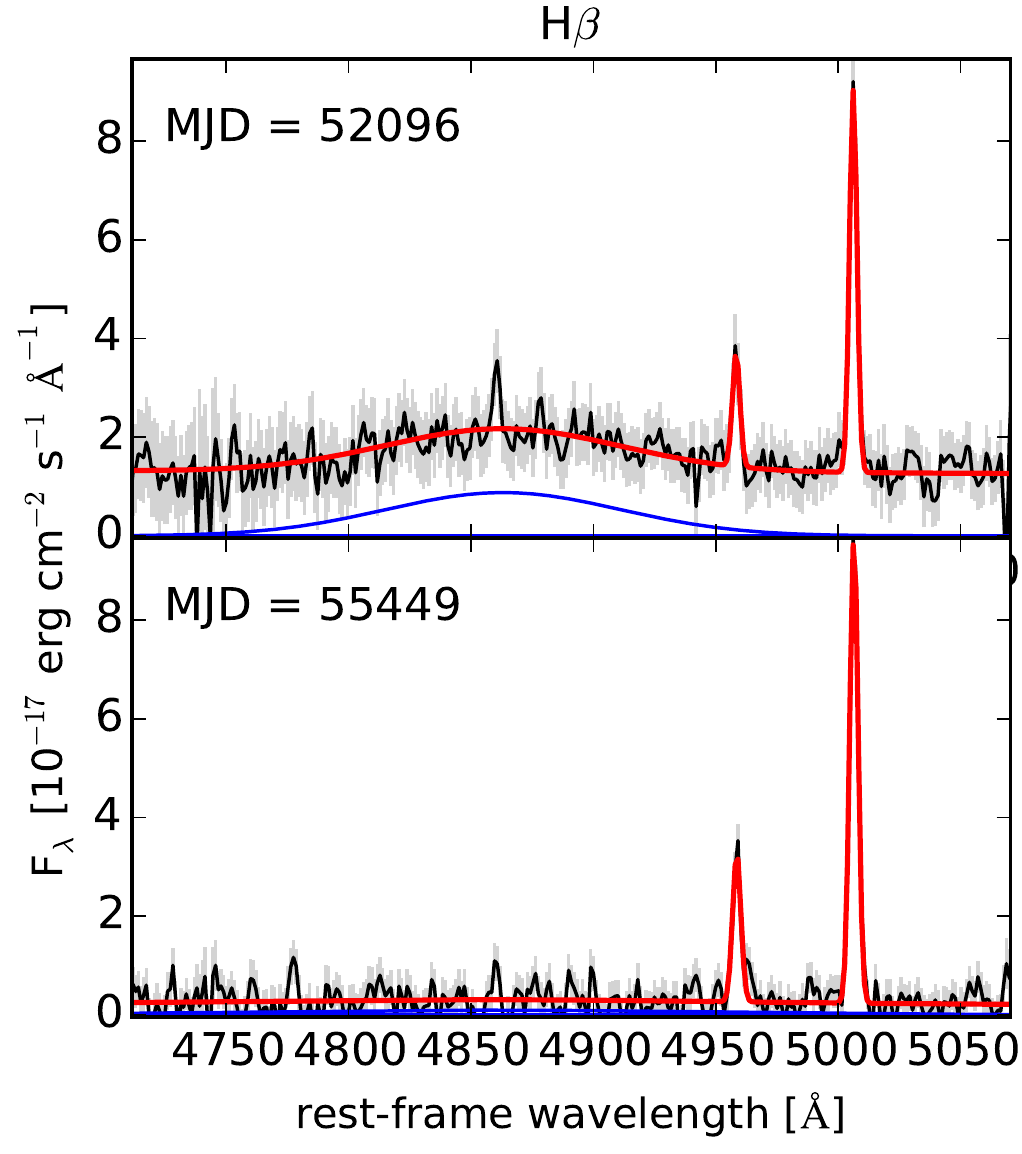} 
\includegraphics[width=0.45\textwidth]{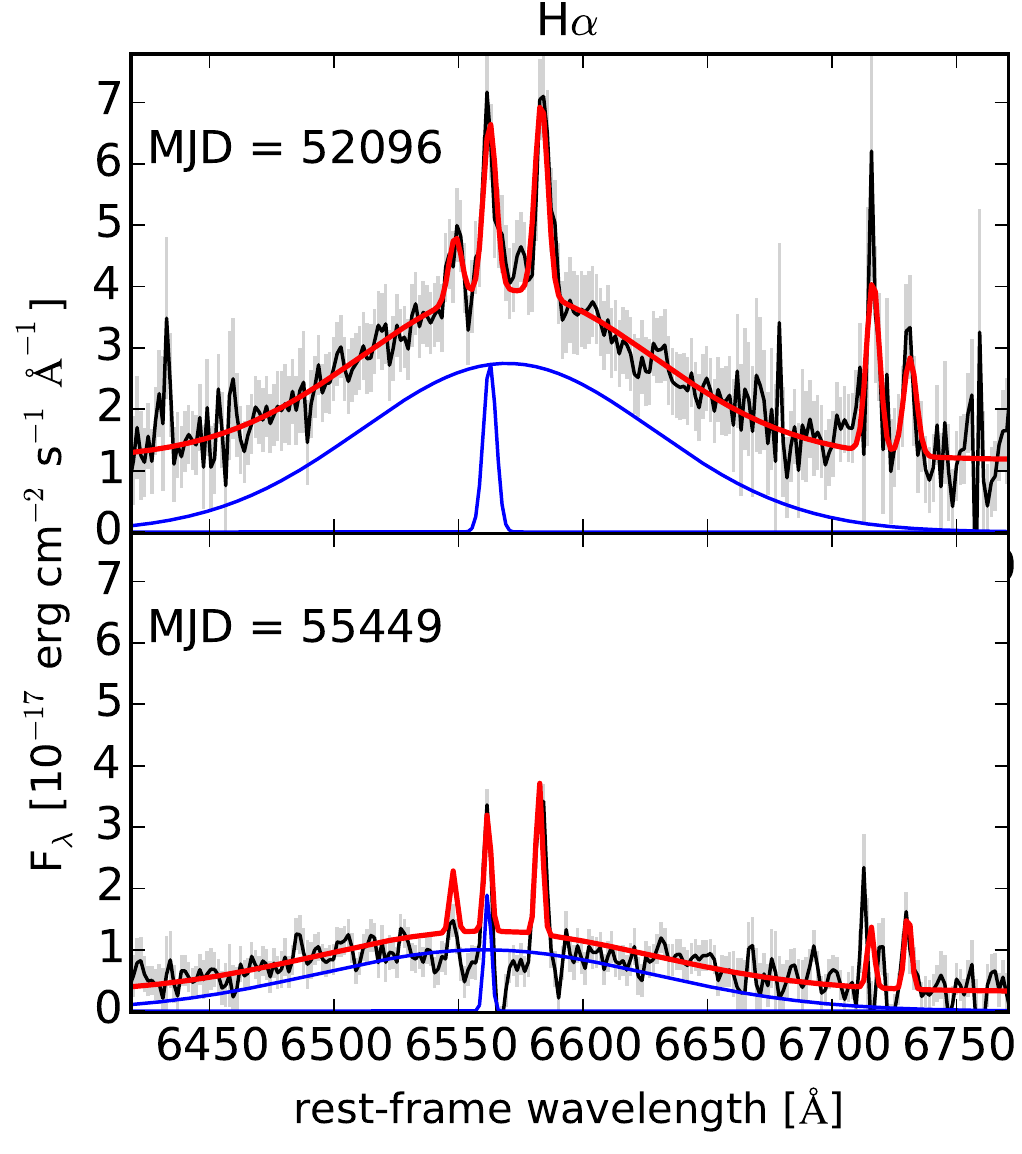} 
\caption{Spectral decomposition for the two epochs of spectra of SDSS J233602.98+001728.7 in the rest-frame, same format as Figure 1. We note that the earlier epoch of spectra at MJD 52096 is a mean stack of four epochs of SDSS spectra taken within a 2-year period (MJD of 51783, 51877, 52199, 52525), during which no strong spectral changes were observed.}
\end{center}
\end{figure*}
	
\section{Spectral Properties of \\ Changing-Look Quasars}

In this section, we describe our decomposition of the multi-epoch spectra of our three CL quasars into host-galaxy and quasar components. We then fit the broad emission lines in the quasar components and analyze their properties. Although single-epoch broadband imaging of our CL quasars is available through SDSS, we do not compare the multi-epoch spectra to their photometry in our analysis. This is primarily because our CL quasars are spatially extended (i.e. resolved) in the imaging, and the SDSS fibers may have different diameters depending on epoch. This causes the stellar contamination in the imaging photometry to vary with aperture, complicating comparisons between the imaging and the multi-epoch spectra. Our approach to investigating the nuclear emission of our CL quasars instead relies on decomposition of the spectra to remove contamination from the host galaxy. We note that the \citet{lamassa15} CL quasar recovered in our search lies in the SDSS Stripe 82 footprint; image differencing light curves of its nuclear emission are presented and discussed in \citet{merloni15}.

\subsection{Spectral Decomposition}
	We decompose the quasar and host-galaxy components in the spectra from both epochs for each of the three CL quasars found in our search. For the spectral decomposition, we follow the general method of \citet{vandenberk06} and \citet{shen15}, with only minor modifications, which is based on fitting quasar and galaxy eigenspectra. Specifically, the CL quasar spectra are each fitted using a mix of quasar and galaxy eigenspectra that are created from a principal component analysis of large samples of SDSS quasar and galaxy spectra. This decomposition method using eigenspectra differs from the approach of \citet{lamassa15}, who instead fit a model of power-law quasar continuum emission, host-galaxy emission generated from a stellar population synthesis model, and emission lines to the multi-epoch spectra. The main advantage of our approach is that the continuum emission is fitted and decomposed empirically without having to rely on the accuracy of a power-law and stellar population synthesis models. For example, a mixture of eigenspectra can more accurately describe a galaxy spectrum consisting of stars with a continuous range of ages in comparison to the simple star formation histories assumed by stellar population synthesis models. The primary disadvantage of our approach is that because narrow emission lines are present in both the quasar and galaxy eigenspectra, separating the narrow line emission from the continuum emission is less straightforward (this does not adversely affect the broad emission lines). Although the CL quasar J0159+0033 recovered in our search was previously discovered and analyzed by \citet{lamassa15}, we nevertheless include it in our analysis below to demonstrate whether our independent spectral decomposition and broad emission line fitting for this object produces results that are consistent with their published values.

	We first correct all spectra in our sample for Galactic extinction, using the maps of \citet{schlafly11} and the Milky Way reddening law of \citet{cardelli89}. To facilitate the spectral decomposition, we resample all our spectra and the eigenspectra to a common wavelength grid of the form log$_{10}$$\lambda = 3.35 + 0.001a$, for integer $a$ from 0 to 5,914. The wavelength coverage of this common grid is wide enough to accommodate all spectra in our sample, and is similar to the native SDSS resolution. For our spectral decomposition, we utilize the eigenspectra from the principal component analysis of $\sim$17,000 SDSS quasar spectra from \citet{yip04a}, and $170,000$ SDSS galaxy spectra from \citet{yip04b}, respectively. Specifically, we fit combinations of the first five quasar and first five galaxy eigenspectra, with their amplitudes (i.e., PCA coefficients) as 10 free parameters. These fits are performed through a simple $\chi^2$ minimization. \citet{yip04a, yip04b} demonstrated that the first five eigenspectra in their PCA analysis captured 98.29\% and 98.37\% of the variance in their quasar and galaxy spectra samples, respectively. Not surprisingly, we find that extending the spectral decomposition to the first 10 quasar and galaxy eigenspectra did not noticeably improve the resulting fits to the observed spectra, and so we only utilize the first five in our analysis for simplicity.
		
	Since the Petrosian radius measured in SDSS imaging for our three CL quasars are between 1.45\arcsec~to 2.58\arcsec, they are spatially extended (i.e. resolved). The host-galaxy and quasar contributions in each resulting SDSS spectrum are thus dependent on the fiber diameter. SDSS-I/II spectra were obtained using 3\arcsec~diameter fibers, while SDSS-III (BOSS) spectra were acquired using 2\arcsec~diameter fibers. For J0159+0033, we decompose the spectra from the two epochs separately because the earlier spectrum was obtained with a 3\arcsec~fiber, while the later spectrum was obtained with a 2\arcsec~fiber. In contrast, both spectra of J0126$-$0839 were obtained with 3\arcsec~fibers, therefore the host-galaxy contribution should be constant between the two spectra. For this object, we decompose both epochs of spectra simultaneously and impose the additional constraint of constant galaxy parameters between the spectra from the two epochs. Finally, for J2336+0017, a total of five epochs of SDSS spectra are available, including four early epochs (with 3\arcsec~fibers) within a 2-year timespan in the observed frame during its quasar-like phase (MJD of 51783, 51877, 52199, 52525), and an epoch 9 years later with a 2\arcsec~fiber. Since the spectral changes in the four early epochs of this object are relatively small, we simply use the mean spectrum of these four epochs in our spectral decomposition to achieve higher signal-to-noise, and our quoted MJD of 52096 is thus actually the mean MJD of the four early epochs. To accommodate the different fiber diameter of the early and later spectra, the galaxy parameters in the fitting are allowed to vary between the mean earlier spectrum and the later spectrum. In Section 4.2, we discuss the evolution in the continuum luminosity of this object over its five separate spectroscopic epochs.

	The results of our spectral decomposition for each of the three CL quasars are shown in Figures 1-3, which display both observed epochs of each CL quasar, as well as their best-fit quasar and galaxy components from eigenspectra. Following previous conventions, we refer to the fitted quasar and galaxy spectra shown in Figures 1-3 as the `reconstructed' quasar and galaxy spectra, while the `decomposed' quasar spectra (not shown) are the original spectra with their corresponding reconstructed galaxy spectra subtracted. All of our broad emission line analysis in Section 4.2 is performed on the decomposed quasar spectra rather than the reconstructed quasar spectra, since this allows us to use the uncertainties on the flux densities from the original spectrum.
	
\subsection{Broad Emission Line Analysis}
	Using our decomposed quasar spectra in both epochs for each of our CL quasars, we measure the properties of the H$\alpha$ and H$\beta$ broad emission lines to estimate black hole masses $M_\mathrm{BH}$ and bolometric Eddington ratios $L_\mathrm{bol}/L_\mathrm{Edd}$. Our broad emission line fitting procedure generally follows the method of \citet{shen11}, in which the single-epoch virial $M_\mathrm{BH}$ estimates are based on the broad emission line Full-Width Half Maximum (FWHMs), as well as a radius-luminosity relation for the broad line region from reverberation mapping of low-redshift AGNs. We emphasize that the narrow emission lines in the decomposed quasar spectra (observed spectrum minus the best-fit galaxy spectrum) we use for the broad emission line fitting are not equivalent to those in the original observed spectrum. This is because narrow emission lines are also present in the galaxy eigenspectra (and thus the best-fit host-galaxy spectrum), which is subtracted to obtain the decomposed quasar spectra. However, we include the narrow emission lines in our analysis below to avoid biases in fitting the broad emission lines.
	
	For the H$\alpha$ region of each object, we use the decomposed quasar spectrum and fit the local continuum emission in the wavelength windows of [6400, 6500]\AA~and [6800, 7000]\AA~to a power-law. In the H$\alpha$ line wavelength window of [6500, 6800]\AA, we fit for the narrow H$\alpha$ component, the [N II] $\lambda\lambda$6548,6584 doublet, and the [S II] $\lambda\lambda$6717,6731 doublet, using a single Gaussian for each emission line. The redshifts of the narrow lines are constrained to be the same, and their widths are constrained to be $<$1200 km s$^{-1}$. The broad component of the H$\alpha$ emission is fit with a Gaussian with width constrained to be $>$1200 km s$^{-1}$, with its central wavelength as a free parameter. 
	
	Similarly, for the H$\beta$ region, we fit a local power-law to the continuum wavelength windows of  [4435, 4700]\AA~and [5100, 5535]\AA, and we fit emission lines in the H$\beta$ line wavelength window  of [4700, 5007]\AA. In the continuum wavelength window, we include the optical Fe II template of \citet{boroson92} in the fit. However, for each of our three CL quasars in the dimmer galaxy-like epoch and two in the brighter quasar-like epoch (J012648.08-083948.0 and J233602.98+001728.7) , the Fe II emission is weak and the template fit is poorly constrained. Thus, the Fe II template is included in the continuum fit only for J0159+0033 in its brighter quasar-like spectral epoch. For the narrow lines, we fit single Gaussians for the narrow component of H$\beta$ and the [O III] $\lambda\lambda$4959,5007 doublet, with widths constrained to be $<$1200 km s$^{-1}$ and redshifts constrained to be the same. The broad H$\beta$ component is fit with a single Gaussian with width $>$1200 km s$^{-1}$, and its central wavelength is a free parameter.
	
	The above spectral fitting produced the FWHM and luminosities of the broad H$\alpha$ and H$\beta$ components, as well as the quasar continuum luminosity at 5100\AA~$\lambda L_{5100}$, for each epoch of our CL quasars. These measured properties of the broad emission lines are tabulated in Table 1, and the spectral fits are also presented in Figures 1-3. All uncertainties are calculated through 10$^3$ Monte Carlo realizations of each spectrum based on their 1$\sigma$ flux density uncertainties. The spectral fitting procedure is performed for each resampled spectrum, and the uncertainties quoted for each parameter are thus the 1$\sigma$ spread in the resulting distributions of resampled parameters.
	
	Using the H$\alpha$ broad emission line FWHM and luminosities from our fits (where detected), single-epoch black hole masses $M_\mathrm{BH, H\alpha}$ are estimated using the relation from \citet{greene10} of 
\begin{align}
M_{\rm BH, H\alpha} = & ~9.7\times10^{6} \left[\frac{{\rm FWHM(H\alpha)}}{1000 {\rm \; km \; s^{-1}}}\right]^{2.06} \nonumber \\
& \times \left[\frac{\lambda L_{\rm 5100}}{10^{44} {\rm \; erg \; s^{-1}}}\right]^{0.519}\; M_\odot,
\end{align}
\noindent based on the radius-luminosity relation of \citet{bentz09}. To verify these H$\alpha$-based estimates, we also estimate single-epoch black hole masses based on our H$\beta$ broad emission line fits $M_\mathrm{BH, H\beta}$ using the relation from \citet{vestergaard06} of 
\begin{align}
M_{\rm BH, H\beta} = & ~10^{6.91} \left[\frac{{\rm FWHM(H\beta)}}{1000 {\rm \; km \; s^{-1}}}\right]^2 \nonumber \\ 	
& \times \left[\frac{\lambda L_{\rm 5100}}{10^{44} {\rm \; erg \; s^{-1}}}\right]^{0.5}\;M_\odot.
\end{align}

	For J0159$+$0033 and J2336$+$0017, the broad H$\beta$ component is not detected in the fainter galaxy-like spectrum. We provide 3$\sigma$ upper limits on the broad H$\beta$ luminosities for these two objects in Table 1. These upper limits are estimated in the fainter galaxy-like epoch by refitting the Gaussian broad H$\beta$ line in 10$^3$ Monte Carlo resamplings of the spectrum, with the FWHM of the broad H$\beta$ component fixed to the width predicted from the observable H$\alpha$ FWHM using Equation 9 from \citet{shen11}.

	For J0126$-$0839, neither broad Balmer emission nor quasar continuum emission are detected in the fainter galaxy-like spectral epoch, and thus the origin of the dimming in this object is poorly-constrained. We provide a 5$\sigma$ upper limit on the continuum luminosity $\lambda L_\mathrm{5100}$ in Table 1, through 10$^3$ Monte Carlo resamplings of the spectrum. To obtain limits for the luminosities of the Balmer emission lines, we must first assume a FWHM. Therefore, we report here the results for two different assumptions: (a) the FWHM in the dim state is the same as that in the bright state, and (b) the FWHM increases as the continuum luminosity drops in order to preserve the black hole mass, as prescribed by Equations 1 and 2 above. Since the continuum is not detected in the dim state, we can only obtain upper limits on the FWHM of H$\alpha$ and H$\beta$ of 7960 and 8920 km s$^{-1}$, respectively. Thus, we assume these (fixed) values of the FWHM in order to obtain upper limits on the line luminosities. The limits we obtain for cases (a) and (b) are given in Table 1. We emphasize that since neither the continuum nor the lines are detected in the dim state, we cannot constrain the dimming mechanism for this object as well as did for the other two objects. Therefore, neither the intrinsic dimming nor the reddening hypothesis can be excluded for J0126$-$0839.

	We estimate the bolometric luminosity $L_\mathrm{bol}$ of our quasars by multiplying the continuum luminosity measured from the decomposed quasar spectra at 5100\AA, $\lambda L_{5100}$, by the bolometric correction factor of 8.1 from \citet{runnoe12}. The Eddington ratio is then $L_\mathrm{bol}$/$L_\mathrm{Edd}$ = $L_\mathrm{bol}$/($1.3\times10^{38}$$M_\mathrm{BH}$), for $M_\mathrm{BH}$ in units of $M_\odot$, and $L_\mathrm{bol}$ in units of erg s$^{-1}$. These inferred quantities are tabulated for both epochs of our CL quasars in Table 2. The $M_\mathrm{BH}$ and $L_\mathrm{bol}$/$L_\mathrm{Edd}$ values derived for each epoch of spectra from H$\alpha$ and H$\beta$ are consistent to within the uncertainties, validating our broad emission line fits.

\begin{figure}[t]
\begin{center}
\includegraphics[width=0.47\textwidth]{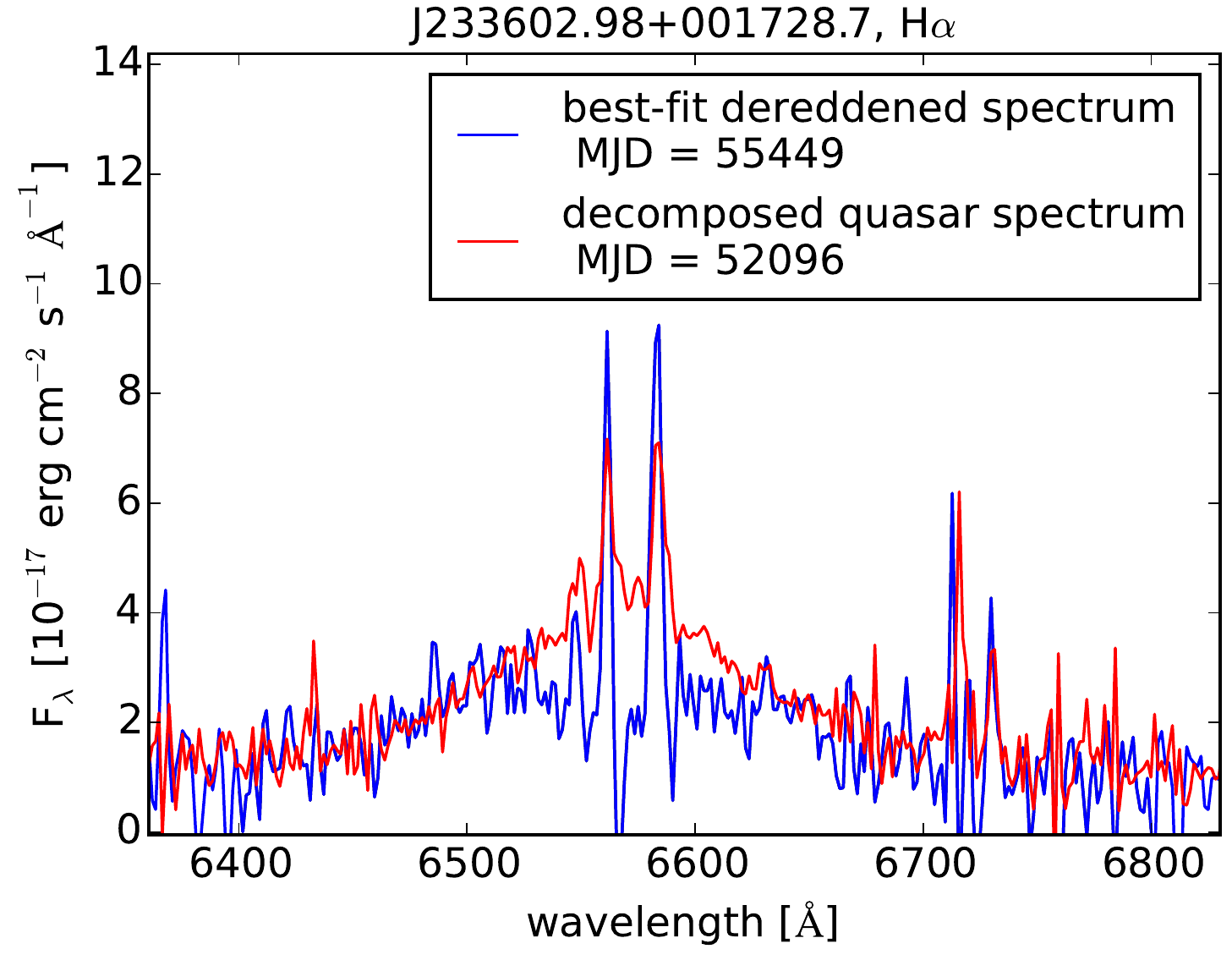} 
\caption{Decomposed quasar spectrum of J2336+0017 in the H$\alpha$ region from the earlier, quasar-like epoch (red), compared to the dereddened decomposed quasar spectrum from the later, galaxy-like epoch (blue). Although changes in dust extinction can reasonably explain the dimming of the continuum emission in this changing-look quasar, the extinction required cannot explain the strong changes in the broad emission line (see discussion in Section 4.1), disfavoring an extinction origin for changing-look quasars.
}
\end{center}
\end{figure}

\section{Discussion}
	In this section, we investigate the origin of the CL quasar phenomenon using our sample. In particular, we will focus on evidence for and against variable dust obscuration, tidal disruption events, Type IIn supernovae, and intrinsic dimming of quasar emission, based on a variety of approaches. We conclude from these investigations that CL quasars are likely to be due to intrinsic dimming of the nuclear emission, as a result of rapidly decreasing accretion rates.

\subsection{A Dust Extinction Origin?}
	We assess whether extinction by an intervening dust cloud along the line of sight outside the broad line region can cause the dimming in the continuum and broad emission lines observed in our CL quasars. Following the method of \citet{lamassa15}, we deredden the decomposed quasar spectrum in the galaxy-like epoch to match the continuum in the quasar-like epoch to assess whether the broad emission line variability can be also be explained by extinction. \citet{lamassa15} reported that while the dimming of the quasar continuum emission in J0159+0033 (recovered in our search) can reasonably be modeled as due to an increase in dust extinction, the changes in the broad emission lines are poorly fit by the same extinction model. While this analysis cannot be performed on J0126$-$0839 since the quasar continuum dimmed completely and is not observed in the galaxy-like spectral epoch, we perform a similar analysis on the CL quasar J2336+0017.
	
	We deredden the decomposed quasar spectrum from the galaxy-like epoch of J2336+0017, and fit an $E(B-V)$ value for which the continuum of the dereddened spectrum best matches the decomposed quasar spectrum from the earlier, quasar-like epoch of this object. This fitted value of $E(B-V) = 0.43$ is determined by minimizing the $\chi^2$ between the two spectra (incorporating all uncertainties) in the wavelength regions outside the H$\alpha$ and H$\beta$ wavelength windows discussed in Section 3.2. A \citet{cardelli89} reddening law for the dust extinction in the host galaxy is assumed, with $R_V = 3.1$. Figure 4 compares the best-fit dereddened quasar spectrum from the later epoch in the H$\alpha$ region to that from the decomposed quasar spectrum from the earlier epoch. Similar to the findings of \citet{lamassa15} for J0159+0033, it is clear that if the dimming of the continuum emission between the two spectral epochs of J2336+0017 is caused purely by dust extinction, the change in extinction required is not consistent with the observed changes in the H$\alpha$ emission. Furthermore, the profile of the broad H$\alpha$ component in Figure 4 broadens between the two epochs of spectra. In an extinction scenario, broadening of the H$\alpha$ profile implies that the emission from the outer, lower-velocity regions of the broad-line region is attenuated more than emission from the inner portions. However, given that the quasar continuum from the central accretion disk is also obscured, such a configuration of the obscuring material is unlikely. These results thus disfavor a dust extinction origin for the CL quasar behavior in J2336+0017.
	
\subsection{Narrow Emission Line Properties}
	\citet{merloni15} suggested that the CL quasar J0159+0033 may be a TDE that was observed serendipitously during and after a luminous flare, producing the quasar-like and galaxy-like spectra, respectively. This scenario is supported by their analysis of the long-term optical light curve of this CL quasar, which appears to show temporal evolution consistent with TDEs. However, they also present several issues with this interpretation stemming from the strong broad and narrow emission lines observed in the SDSS spectrum. Specifically, the mass of gas in the broad line region inferred by \citet{merloni15} from the spectrum (of order $\gtrsim$100 $M_\odot$) is significantly more than could be provided from TDE debris. Furthermore, the possible TDE in this galaxy is unlikely to have ionized the gas producing the observed narrow lines, since the light travel time to the narrow line region (distances of kpc scales, e.g. \citealt{liu13, hainline13}) is $\sim$10$^{3-4}$ years. Interestingly, the narrow line ratios of this CL quasar are not consistent with stellar photoionization, leaving a long-lived AGN as the most plausible power source. Here, we examine the narrow line properties for the three CL quasars in our sample, and compare them to those from AGN and TDE spectra.
	
\begin{figure}[t]
\begin{center}
\includegraphics[width=0.47\textwidth]{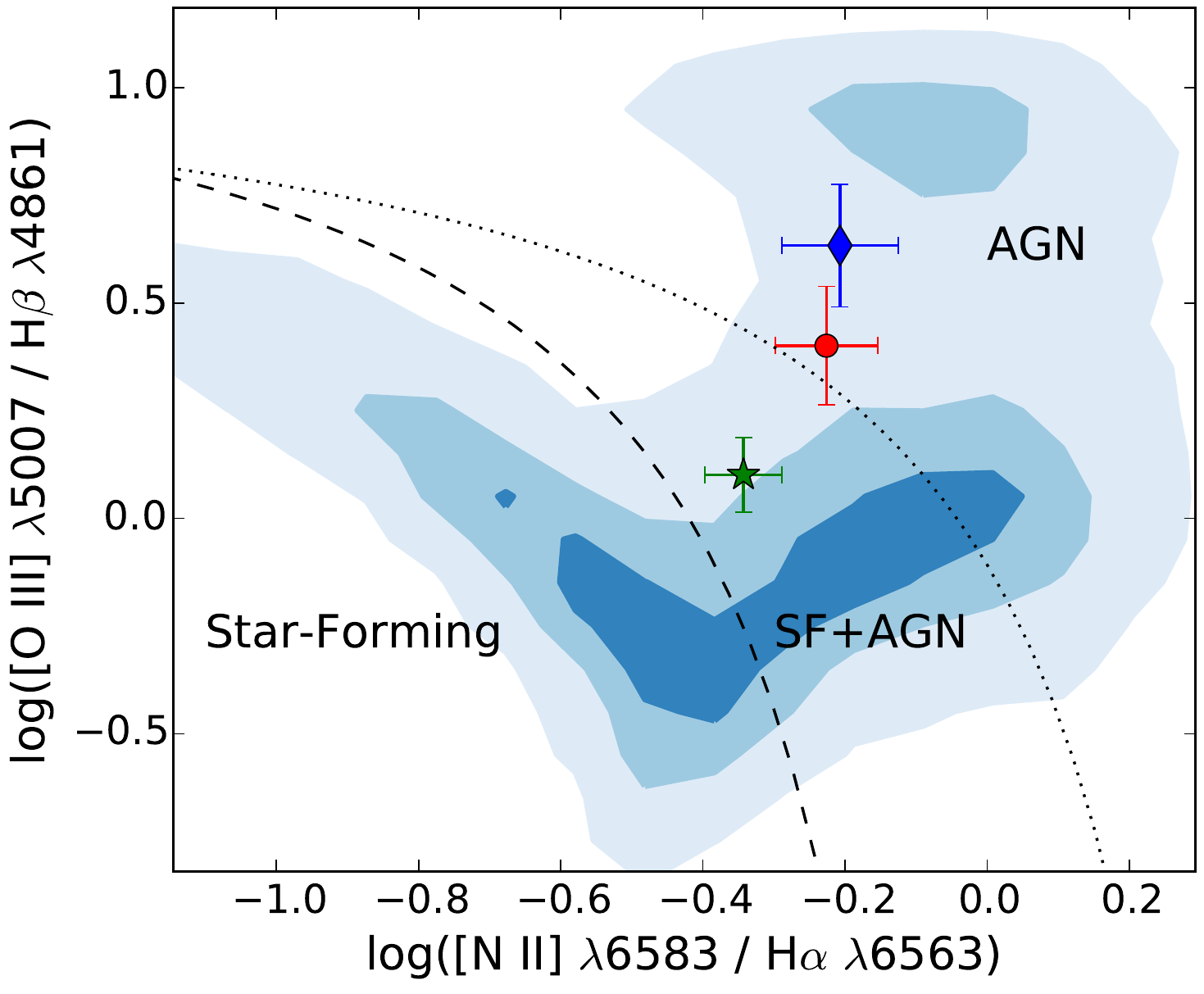} 
\caption{BPT diagram of the three changing-look quasars in our sample, based on the emission line ratios measured in their latest SDSS spectrum. The line ratios for SDSS J015957.64+003310.5 (red circle), SDSS J012648.08-083948.0 (green star), SDSS J233602.98+001728.7 (blue diamond), and their 1$\sigma$ uncertainties are shown along with all emission line galaxies in SDSS-III DR12 (blue contours) for comparison. The BPT diagram classification schemes of \citet{kauffmann03} (dashed line) and \citet{kewley01} (dotted line) are shown. The changing-look quasars appear to exhibit emission line ratios 
that are consistent with AGN-like or composite (AGN and stellar) ionizing continuum emission rather than powered purely by star formation alone.
}
\end{center}
\end{figure}
			
	Our eigenspectra-based spectral decomposition method in Section 3.1 does not allow a clean separation of the narrow emission lines from the underlying stellar and quasar continuum emission. We therefore utilize the stellar population and emission line fits to SDSS DR12 spectra by \citet{thomas13}\footnote{www.sdss.org/dr12/spectro/galaxy\_portsmouth/\#kinematics} to examine our CL quasars on a classic BPT \citep{baldwin81} diagram. These fits are performed using the Gas and Absorption Line Fitting (GANDALF) software described in \citet{sarzi06}, and include Gaussian fits to a variety of emission lines to obtain line fluxes. We specifically use the [N II] $\lambda$6583 to H$\alpha$ and [O III] $\lambda$5007 to H$\beta$ narrow line ratios measured for the latest SDSS spectrum of our CL quasars (during the fainter, galaxy-like state), as shown in the BPT diagram in Figure 5. Figure 5 also indicates these line ratios for all galaxies in SDSS-III DR12 that have all four emission lines detected at $>$3$\sigma$ significance, as well as the BPT classification scheme of \citet{kauffmann03} and \citet{kewley01}, which distinguishes galaxies with emission lines ionized by an AGN-like, stellar-like, and composite (AGN and stellar) continuum. These line ratios for J0159+0033 published by \citet{thomas13} are consistent with those independently determined by \citet{lamassa15} and \citet{merloni15} for the latest SDSS spectrum. Figure 5 demonstrates that the emission lines in our sample of CL quasars are ionized at least in part by an AGN-like continuum, disfavoring a TDE scenario. For J0126$-$0839, the line ratios appear to be composite, which may be due to strong star-formation in its host galaxy.
		
	We also examine the fluxes of the narrow emission lines, focusing on the strongest narrow line, [O III] $\lambda$5007. Line fluxes reported for [O III] $\lambda$5007 from \citet{thomas13} for the three CL quasars in our sample range from (1.2 - 1.7) $\times$ 10$^{-15}$ erg s$^{-1}$ cm$^{-2}$, which correspond to luminosities of (1.7 - 5.3) $\times$ 10$^{41}$ erg s$^{-1}$ in the rest-frame. This is in strong contrast to narrow line emission observed in spectra of UV/optical TDEs, which show no or significantly fainter [O III] $\lambda$5007 emission \citep[e.g.][]{gezari06, gezari09, gezari12, holoien14, chornock14}, further disfavoring a TDE explanation for CL quasars. The faint [O III] $\lambda$5007 lines detected in these TDEs also have ratios relative to other lines that are consistent with star formation rather than AGN photoionization; this may not be surprising given the long light-travel times (10$^{2-3}$ years) for the ionizing TDE continuum to reach the narrow line region gas. Thus, the combination of timescale arguments, narrow emission line luminosities, and emission line ratios in our sample of CL quasars all suggest that CL quasars are linked to quasar activity rather than TDEs.
	
\begin{figure}[t]
\begin{center}
\includegraphics[width=0.48\textwidth]{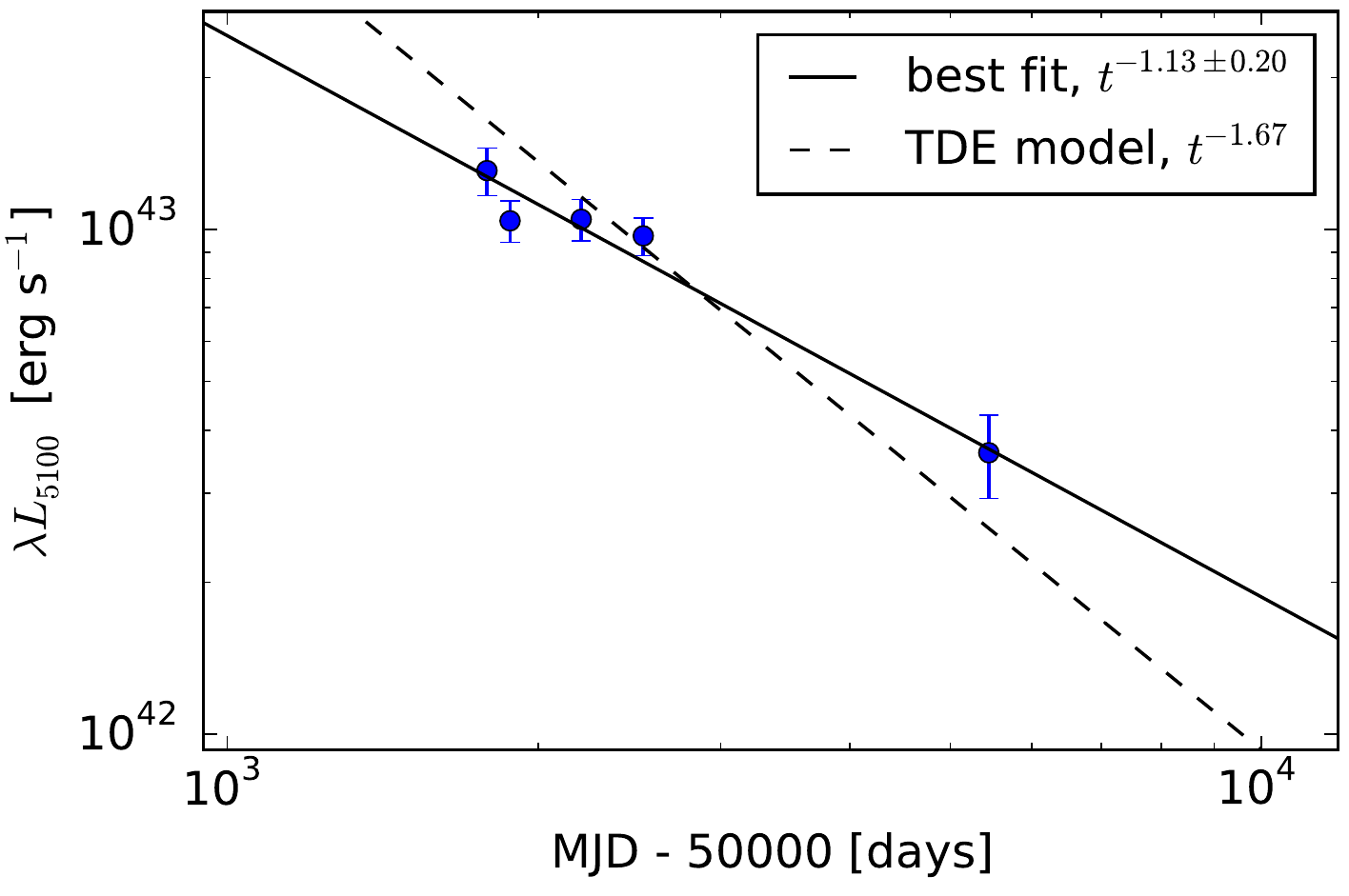} 
\caption{The light curve of the continuum luminosity in the decomposed quasar spectra of SDSS J233602.98+001728.7. The best-fit power-law model is shown as a black solid line, and the best-fit power-law model with spectral index fixed to the $-5/3$ value expected from tidal disruption events is shown as a black dashed line.}
\end{center}
\end{figure}
	
\subsection{The Nuclear Light Curve of J2336+0017}
	The CL quasar J2336+0017 in our sample has a total of five epochs of spectra, including four early epochs in its quasar-like state over approximately 2 years in the observed frame, and one later epoch in its galaxy-like state. Although we stacked the four early epochs of spectra in our earlier spectral analysis since the broad emission lines did not show noticeable evolution, here we decompose separately all five epochs of spectra to study the time-evolution of the quasar continuum emission. A $t^{-5/3}$ temporal evolution of the decaying continuum luminosity is often taken to be an observational signature of TDEs, since it is the theoretical rate at which the tidal debris is expected to fall back towards the SMBH \citep{rees88, lodato09, strubbe09}, although the luminosity evolution may not necessarily follow this rate \citep{lodato11}. To test the TDE explanation further, we compare the quasar light curve from the five epochs of spectra to the expected TDE luminosity evolution.
	
	Using the same decomposition method as described in Section 3.1, we measure the continuum luminosity at 5100\AA~$\lambda L_{5100}$ of the decomposed quasar spectrum for the five spectroscopic epochs of J2336+00172. Figure 6 presents the light curve, along with the best-fit power-law of $t^{-1.13\pm0.20}$, and the best-fit TDE $t^{-5/3}$ model. The uncertainty on this power-law index is calculated by 10$^3$ Monte Carlo resamplings of each point on the light curve from a Gaussian based on their 1$\sigma$ uncertainties, to produce 10$^3$ resampled light curves. Each resampled light curve is refit, and the quoted 1$\sigma$ uncertainty on the power-law index is the 1$\sigma$ spread from the resampled light curves. The light curve's best-fit power-law index of $-1.13\pm0.20$ is statically shallower than the $-5/3$ model predicted for TDEs, although this only weakly disfavors the TDE model due to the sparse light curve sampling and uncertainties in the models for temporal evolution of TDE luminosity. 

\subsection{A Type IIn Supernovae Origin?}
	Type IIn supernovae (SNe) frequently display broad H Balmer emission and blue UV continuum emission in their optical/UV spectra, which can appear qualitatively similar to AGN spectra during certain phases of their evolution \citep{filippenko89}. Thus, it may be possible that our CL quasars are actually serendipitously discoveries of Type IIn SNe. This possibility is ruled out for two of our CL quasars (J0159$+$0033 and J2336$+$0017) because broad H$\alpha$ emission is still detected in their later galaxy-like spectra approximately 7 years later in the rest-frame. Furthermore, the narrow line ratios for these two objects are clearly AGN-like (see Figure 5), indicating the presence of nuclear activity. For J0126$-$0839, no broad emission lines or quasar continuum emission is detected in the later galaxy-like spectral epoch, leaving this dimming timescale poorly constrained. Furthermore, the narrow line ratios for this object fall in the `composite' rather than the AGN region of the relevant diagnostic line ratio diagrams, making a Type IIn SN explanation for this object somewhat more plausible at first glance. However, the H$\alpha$ luminosity of J0126-0839 in the bright state (10$^{42.0}$ erg s$^{-1}$ from Table 1) is approximately 1-2 orders of magnitude more than the peak H$\alpha$ luminosities normally observed in Type IIn supernovae \citep[e.g. see Figure 14 of ][]{taddia13}. The strong H$\alpha$ emission observed in Type IIn supernovae emerges several months after their light curve peaks, when the continuum emission has dimmed by several magnitudes and is red \citep{filippenko97}. In contrast, the strong H$\alpha$ emission observed in the bright-state spectrum of J0126-0839 is coincident with highly-luminous and blue continuum emission. P Cygni profiles and other asymmetries often observed in the H$\alpha$ line of Type IIn SNe are also not seen in our CL quasars. For these reasons, we disfavor the Type IIn supernovae scenario for J0126$-$0839, but this possibility cannot be conclusively ruled out because the nature of our search for CL quasars using a large data set may preferentially uncover such rare phenomena. Follow-up X-ray and radio observations of the nucleus of J0126$-$0839 may be useful to confirm its AGN nature.
	
\subsection{Infall Timescales}
	Our investigations above generally disfavor TDEs, Type IIn SNe, and dust obscuration as the origin of CL quasars, and instead suggest that this phenomenon is due to intrinsic dimming of the quasar emission from rapidly decreasing accretion rates. Here, we assess whether the $\lesssim$10 year transition timescales we observe for our sample of CL quasars are consistent with the infall timescale of gas in the radiation-pressure dominated inner regions of Shakura-Sunyaev \citep{shakura73} thin accretion disks; this is also the timescale on which changes in the accretion rate are reflected in changes in the continuum luminosity. Using Equation 5 from \citet{lamassa15}, we find that the infall timescale for our sample of changing look quasars are approximately 42, 38, and 868 years for J0159+0033, J0126$-$0839, and J2336+0017, respectively. Although these infall timescales are longer than the transition timescales we observed, we note that the transition timescales observed in our CL quasars are lower limits since the multi-epoch spectra do not encompass the full transition (e.g., it is likely that the CL quasars have begun dimming before the first spectral epoch). Furthermore, \citet{lamassa15} also point out that magneto-hydrodynamic simulations of quasar accretion flows have suggested that the infall timescale may be a factor of a few shorter than these analytical estimates \citep{krolik05}, and much closer to the observed $\sim$10 year timescales. However, the 868 year infall timescale estimated for J2336+0017 is problematic in this interpretation, and may indicate that other processes such as thermal or dynamical instabilities may be present in the accretion flows of CL quasars.
	
\section{Conclusions}
	The discovery of CL quasars presents a new opportunity to study the nuclear environment and structure of quasars, once the origin of this phenomenon is understood. To provide a larger sample of these objects, we performed an archival spectroscopic search in SDSS, yielding three CL quasars, including two new cases. Using this sample, we investigate the detailed properties of their quasar continuum emission and broad and narrow emission lines, with the goal of attempting to discriminate between various possibilities for the origin of this phenomenon. The primary results of our investigation can be summarized as below:
	
\begin{itemize}
\item{The three CL quasars in our sample appear to show similar properties: they are luminous ($L_\mathrm{bol} \sim 10^{44.0 - 44.5}$ erg s$^{-1}$) quasars at relatively low redshifts ($z \sim 0.2 - 0.3$) that display strong dimming of the quasar continuum and the broad H Balmer emission lines over timescales of approximately 5 to 7 years in the rest-frame.} 
\item{ In the two CL quasars for which the broad emission lines are detectable in both spectral epochs, the decrease in broad line luminosity coincides with broadening of the broad line widths, such that the derived black hole masses are preserved. This is consistent with a rapid change in their Eddington ratios, which decrease} from $L_\mathrm{bol}/L_\mathrm{Edd} \sim0.03-0.005$ until the broad H Balmer lines have dimmed significantly or disappeared. In one CL quasar (J0126-0839), all quasar continuum emission and broad emission lines have disappeared below the SDSS detection limit.
\item{Changes in dust extinction required to match the dimming in the quasar continuum cannot account for the changes in the broad emission lines in either of our two CL quasars for which this analysis was possible, disfavoring an extinction origin for this phenomenon. Narrow emission line diagnostics show that our CL quasars all have luminous narrow lines with line ratios consistent with at least partially AGN-like ionizing emission. We argue that these narrow line properties favor a scenario in which the quasar continuum dims intrinsically over a TDE origin for this phenomenon.}
\end{itemize}

	If the intrinsic dimming of the quasar emission favored by our analysis is due to draining of the underlying quasar accretion disk, the CL quasar phenomenon will provide a unique new laboratory to study the accretion flow and nuclear environment in luminous AGNs. Long-term spectroscopic and multi-wavelength monitoring of the currently-known changing-look quasars can help further elucidate the origin of CL quasar transitions. For example, observations of the return of any CL quasar to a bright quasar-like state would provide additional constraints on the physical mechanism of this phenomenon as well as estimates of its duty cycle. Current multi-object spectroscopic programs and time-domain imaging surveys are well-poised to serendipitously discover many more CL quasars. For example, the Time-Domain Spectroscopic Survey \citep{morganson15} in SDSS-IV will provide repeat spectroscopy of several thousand low-redshift quasars, while the Pan-STARRS 3$\pi$ survey has repeated imaged 30,000 deg$^2$ of sky (including the SDSS imaging footprint). With future instruments and surveys such as the Dark Energy Spectroscopic Instrument \citep[DESI,][]{levi13} and the Large Synoptic Survey Telescope \citep[LSST][]{ivezic08}, discovery of such rare phenomena will become more routine.
	
\acknowledgments
JJR thanks James R.A. Davenport for helpful discussions. JJR acknowledges support provided by 
NASA through Fermi Guest Investigator grant NNX14AQ23G.

Funding for SDSS-III has been provided by the Alfred P. Sloan Foundation, the Participating Institutions, the National Science Foundation, and the U.S. Department of Energy Office of Science. The SDSS-III web site is http://www.sdss3.org/.

SDSS-III is managed by the Astrophysical Research Consortium for the Participating Institutions of the SDSS-III Collaboration including the University of Arizona, the Brazilian Participation Group, Brookhaven National Laboratory, Carnegie Mellon University, University of Florida, the French Participation Group, the German Participation Group, Harvard University, the Instituto de Astrofisica de Canarias, the Michigan State/Notre Dame/JINA Participation Group, Johns Hopkins University, Lawrence Berkeley National Laboratory, Max Planck Institute for Astrophysics, Max Planck Institute for Extraterrestrial Physics, New Mexico State University, New York University, Ohio State University, Pennsylvania State University, University of Portsmouth, Princeton University, the Spanish Participation Group, University of Tokyo, University of Utah, Vanderbilt University, University of Virginia, University of Washington, and Yale University.

Based on observations obtained with the Apache Point Observatory 3.5-meter telescope, which is owned and operated by the Astrophysical Research Consortium.

\bibliography{bibref}
\bibliographystyle{apj}

\begin{appendix}

\begin{figure*}[t]
\begin{center}
\includegraphics[width=0.7\textwidth]{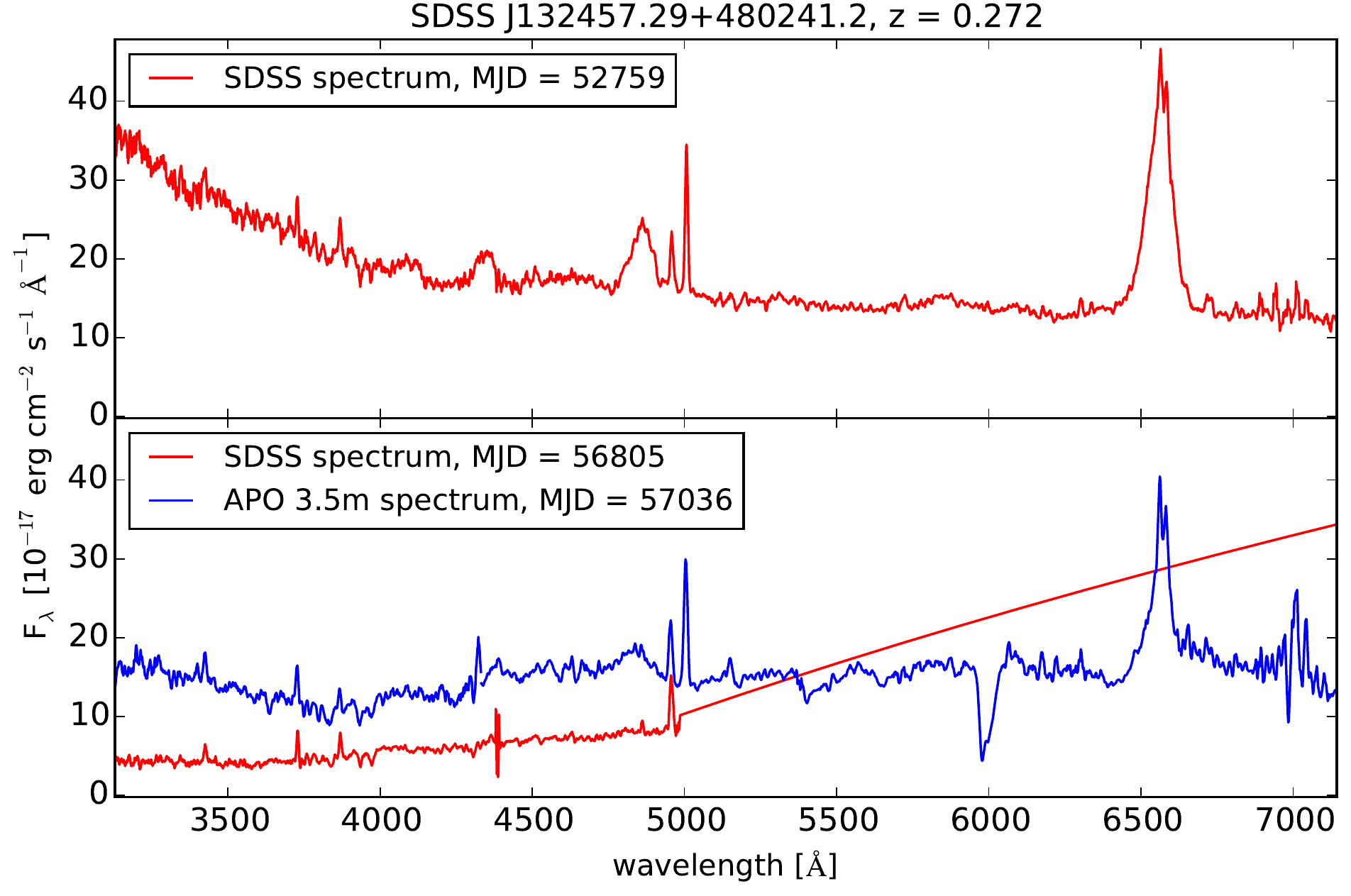} 
\caption{SDSS and APO 3.5m rest-frame spectra of SDSS J132457.29$+$480241.2. The straight red line in the SDSS spectrum of this object at MJD = 56805 is directly from the SDSS spectrum file, and indicates that this spectrum was not obtained or reduced correctly.}
\end{center}
\end{figure*}

\section{A. The CL Quasar Candidate SDSS J132457.29$+$480241.2}

	In addition to the three CL quasars yielded by our systematic search, we identified one additional possible CL quasar in our visual inspection (SDSS J132457.29$+$480241.2) that is classified as {\tt CLASS = `QSO'} in its earlier SDSS spectrum, and {\tt CLASS = `GALAXY'} in its latest SDSS spectrum (shown in Figure 7). However, the red side of the latest SDSS spectrum is corrupted beyond 5000\AA, although the blue side appears to show the disappearance of broad H$\beta$ and H$\gamma$ emission, along with dimming of the quasar continuum, similar to CL quasars. The fiber (527) of the latest SDSS spectrum is known to be affected by columns of bad pixels in the red camera of the BOSS spectrograph, which is likely to be the cause of the corrupted spectrum (and affects spectra from neighboring fibers on this and other plates). 
	
	Nevertheless, to verify whether this object is indeed a CL quasar, we obtained additional optical long-slit spectra using the Dual Imaging Spectrograph on the Astrophysical Research Consortium 3.5m telescope at Apache Point Observatory, with wavelength coverage of $\lambda \sim 3400 - 9200$ \AA. Three 15 minute exposures where taken on 14 January 2015 UT, at a spectral resolution of R$\sim$800 using the B400/R300 grating settings, and a 1.5$\arcsec$ slit. The seeing was 1.8$\arcsec$ on this night, and the observations were obtained at airmass of approximately 1.04. Spectra of the spectrophotometric standard star Feige 34 were also obtained for flux-calibration and removal of atmospheric absorption, and HeNeAr lamps were used to obtain a wavelength solution. These spectra were bias and flat-field corrected, wavelength- and flux-calibrated, and corrected for atmospheric extinction using standard IRAF procedures. 
	
	The calibrated APO 3.5m spectrum is displayed in the lower panel of Figure 7, which shows that although the quasar continuum, broad H$\alpha$, and broad H$\beta$ emission has dimmed, they remain prominent, and thus we do not include this object in our CL quasar sample. Although the disagreement between the APO 3.5m spectrum and the latest SDSS spectrum is likely due to the known column of bad pixels, a scenario in which this object is a CL quasar that dimmed in the latest SDSS spectrum to a galaxy-like state, then rebrightened back to a quasar-like state in the APO 3.5m spectrum cannot be ruled out. Given the short period between the latest SDSS spectrum and APO spectrum (182 days in the rest-frame), and the known issues with the latest SDSS spectrum in the red side, rebrightening of a CL quasar appears unlikely, but additional spectral monitoring may be warranted.
	
\section{B. Rejected CL Quasar Candidates}

Our systematic search for CL quasars based on multi-epoch SDSS spectra (outlined in Section 2.1) yielded 117 candidates for which the SDSS spectral pipeline classification switched between {\tt CLASS = `GALAXY'} and {\tt CLASS = `QSO'} (or vice versa) in repeat spectra. Visual inspection of these 117 candidates resulted in the 3 confident CL quasar candidates, while the remaining 114 were rejected because they did not display the disappearance of any broad line and continuum emission characteristic of CL quasars. Figure 8 shows example repeat spectra of some of these candidates that were rejected in our visual inspection, which includes 24 objects in which broad H$\alpha$ emission was outside the spectral range in one spectral epoch but not the other. Figure 8 also shows the repeat spectra of one object (SDSS J032332.83$-$000740.3) for which the fiber was positionally offset between the two spectral epochs. A positional offset of the fiber (e.g. off the nuclear region) can artificially cause changing-look behavior, although this would also cause significant artificial variability in the host-galaxy emission.

\begin{figure*}[t]
\begin{center}
\includegraphics[width=0.49\textwidth]{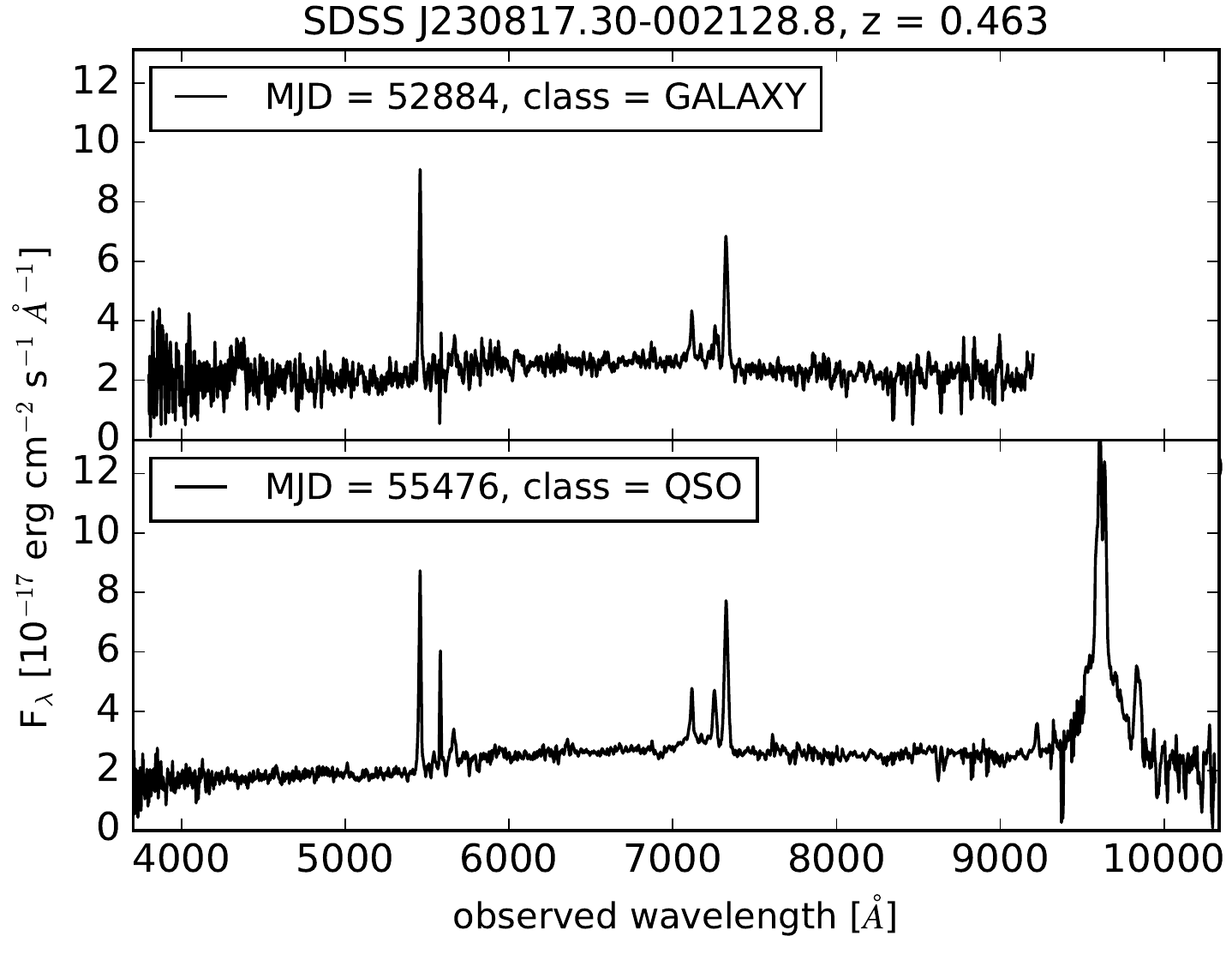} 
\includegraphics[width=0.49\textwidth]{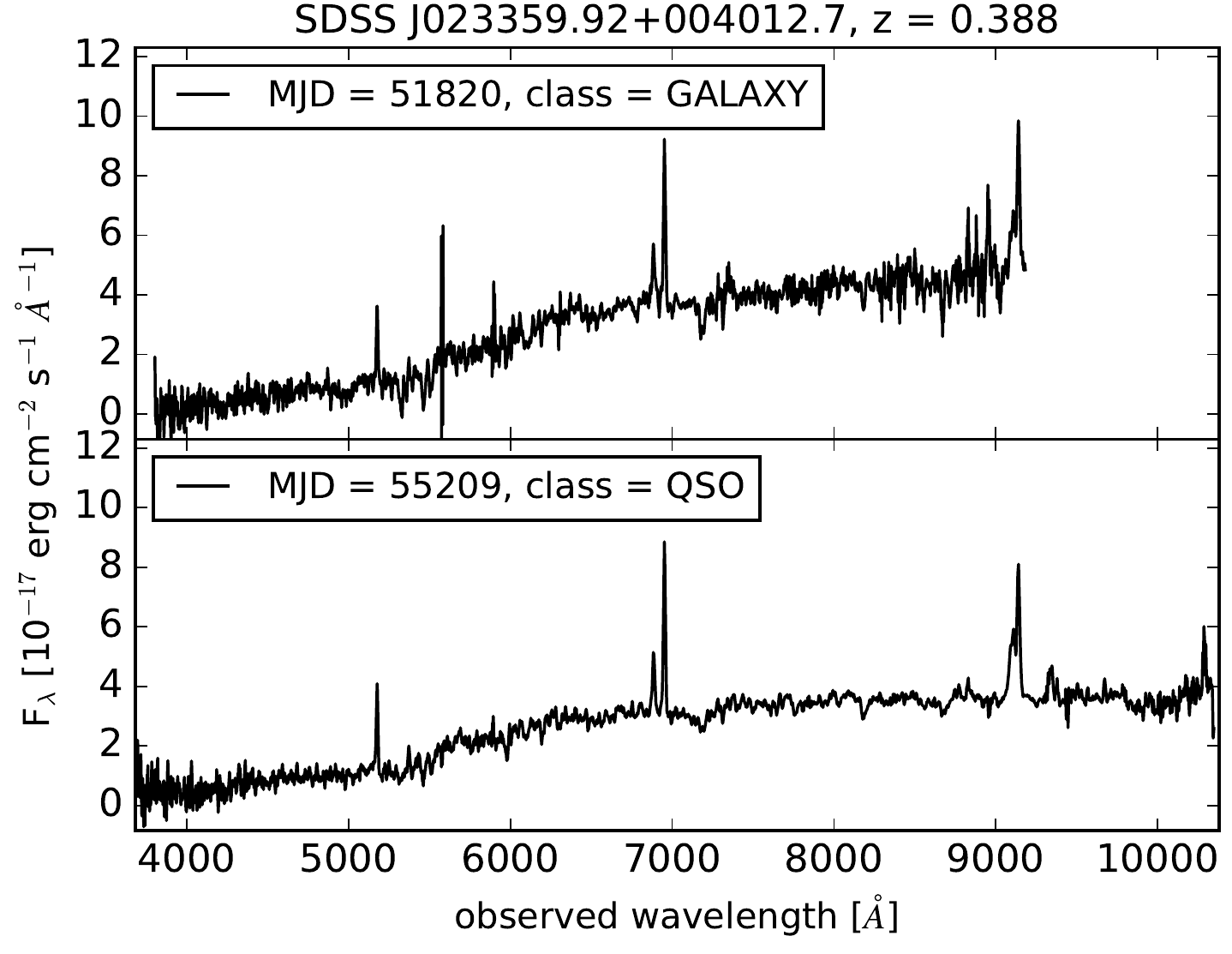}  \\
\includegraphics[width=0.49\textwidth]{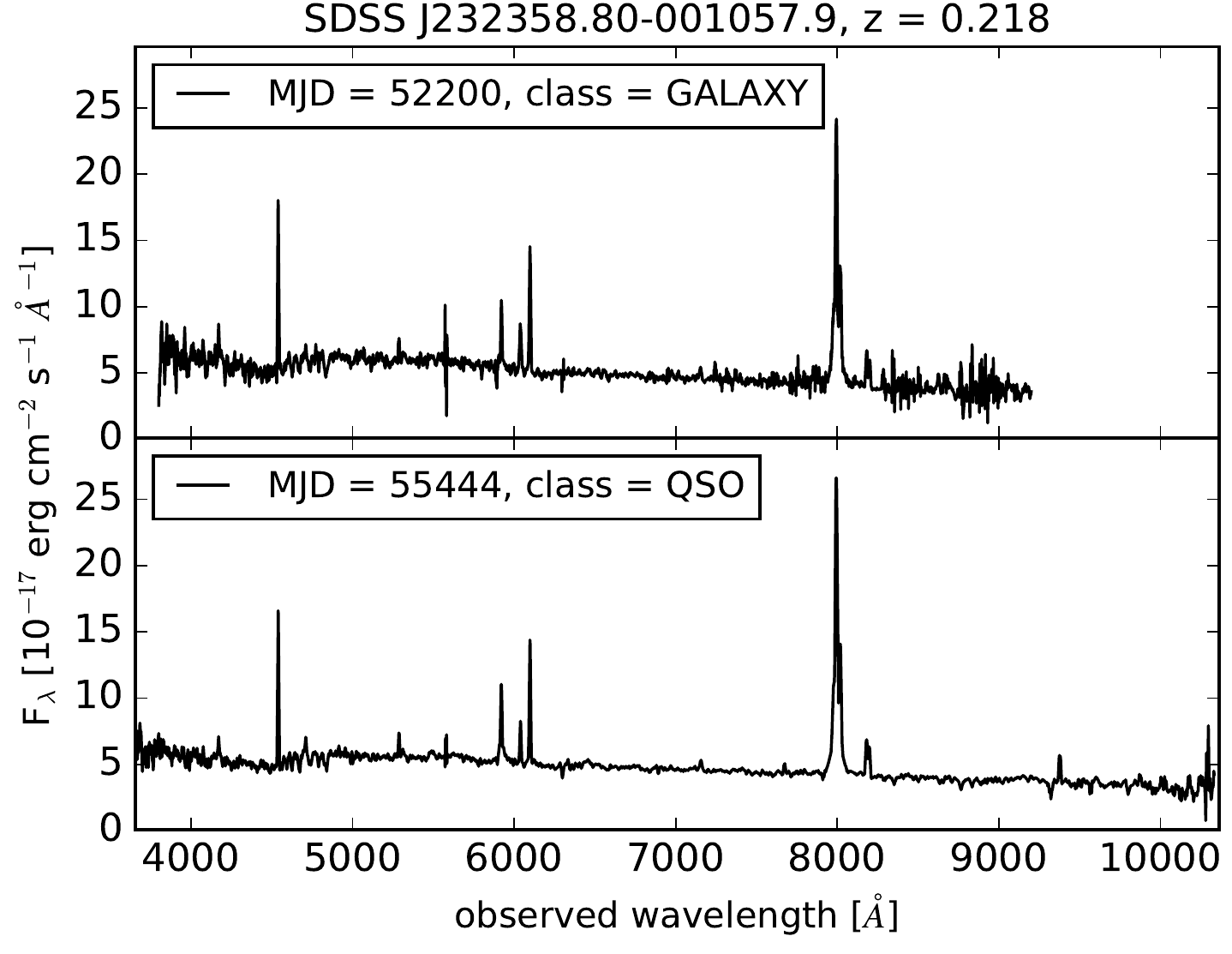} 
\includegraphics[width=0.49\textwidth]{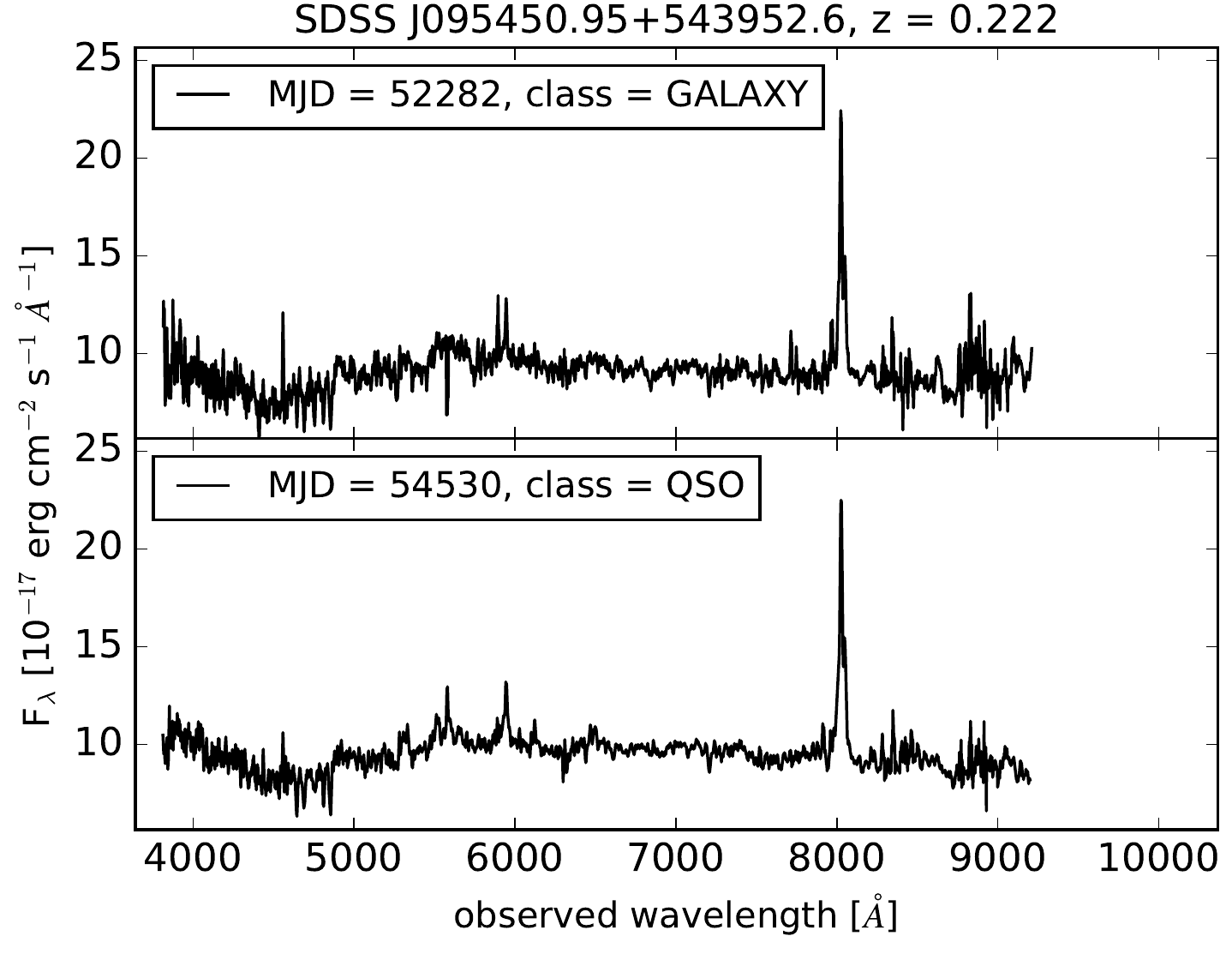} \\
\includegraphics[width=0.49\textwidth]{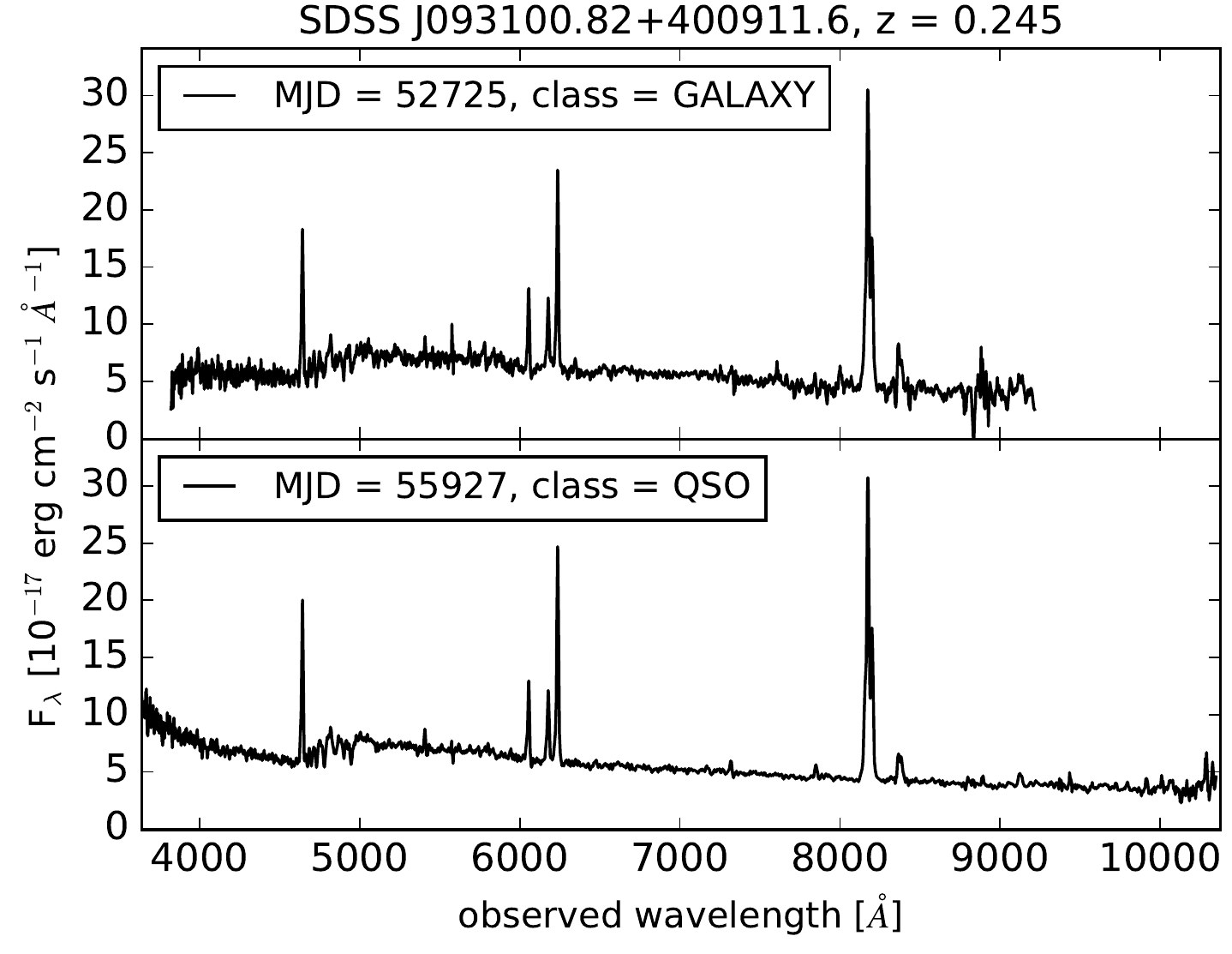} 
\includegraphics[width=0.49\textwidth]{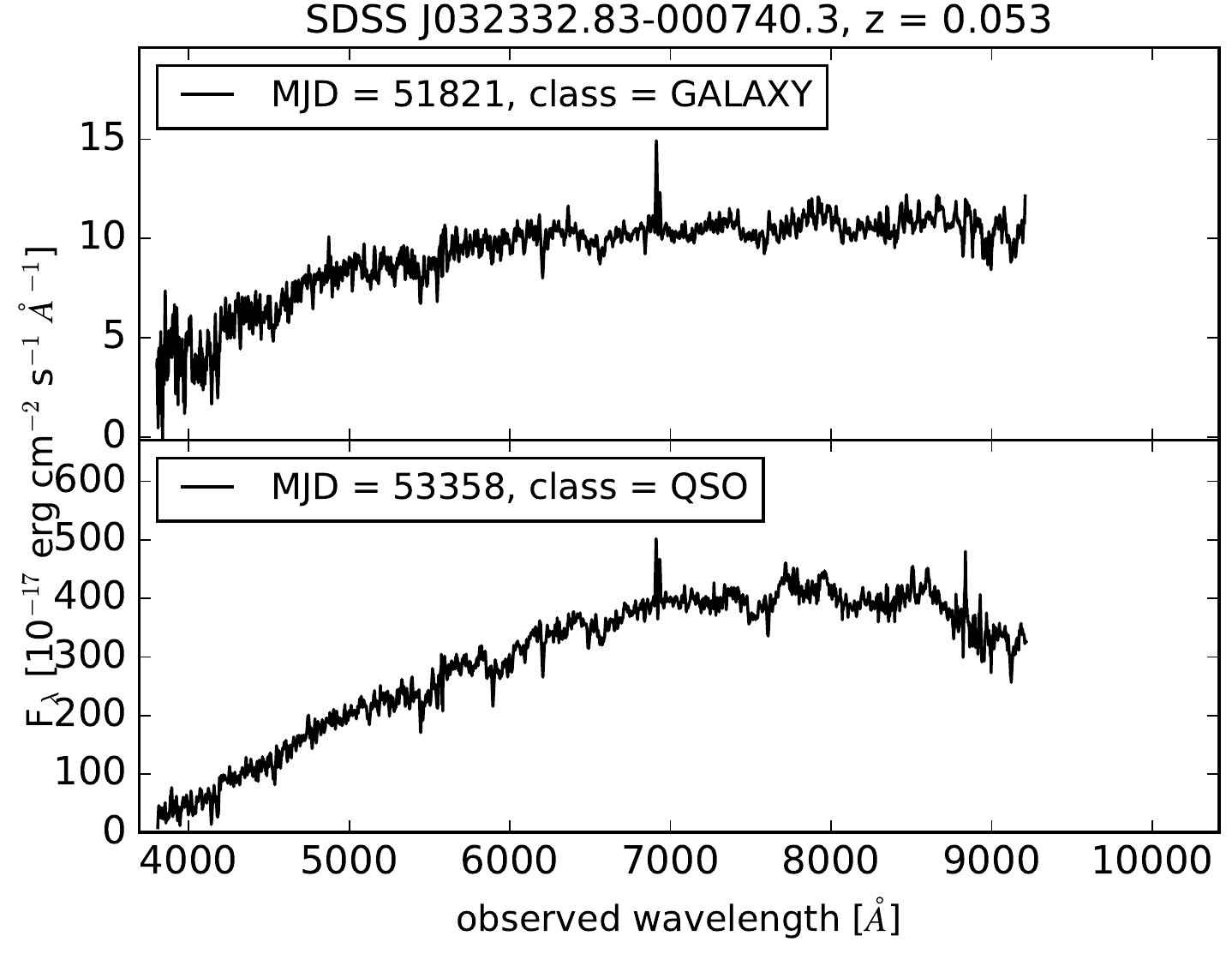} 
\caption{Example repeat SDSS spectra of the CL quasar candidates that were rejected in our visual inspection. The MJD and SDSS pipeline classification of each spectral epoch are shown. These rejected candidates include objects in which the broad H$\alpha$ emission is outside the spectral range in one epoch but not the other (e.g., SDSS J230817.30$-$002128.8 and SDSS J023359.92$+$004012.7). For one object (SDSS J032332.83$-$000740.3), the fiber was positionally offset, causing the observed artificial variability in the host-galaxy emission. Aside from this object and the three CL quasars, all other candidates were rejected because our visual inspection of their repeat spectra did not reveal dramatic disappearance or appearance of broad emission lines.}
\end{center}
\end{figure*}

\section{C. Transition Types}

	Our search criteria for CL quasars should allow us to find cases in which objects spectroscopically transformed from quasar-like to galaxy-like, as well as the reverse. However, in our visual inspection of the CL quasar candidates, we find no confident cases of galaxy-like to quasar-like CL quasars. Although this reverse transition (galaxy-like to quasar-like) has been observed previously in Seyfert galaxies \citep[e.g.,][]{goodrich95, shappee14}, these Seyferts are at lower redshifts and lower luminosities than our CL quasars. It is thus unclear whether the lack of reverse transitions in our sample is due to selection effects, a random result stemming from our small sample size, or a true feature of CL quasars. 
	
	We first consider potential selection effects that may cause us to select CL quasars of only the quasar-like to galaxy-like variety. For example, if SDSS systematically targeted quasars for repeat spectroscopy in greater numbers than galaxies, then naturally we would find more quasar-like to galaxy-like transitions. To test this possibility, we identify the number of repeat spectra in SDSS DR12 of objects with {\tt CLASS = `GALAXY'} and objects with {\tt CLASS = `QSO'}, using the same search criteria described in Section 2.1. There are 11,438 repeat spectra of 8,865 unique galaxies, and 7,109 repeat spectra of 5,990 quasars (some objects had more than two epochs of spectra). This test shows that our sample of only quasar-like to galaxy-like CL quasars is not simply due to a lack of repeat galaxy spectra in SDSS. However, the myriad other potential selection effects, such as those stemming from the disparate SDSS galaxy and quasar completeness magnitude limits, the quasar and galaxy targeting methods in SDSS, and differences in the galaxy populations between those that host quasars and those targeted by SDSS spectroscopy make it difficult to draw robust conclusions on selection effects without more careful considerations.
		
	We next consider whether our yield of three CL quasars of the quasar-like to galaxy-like variety is simply a random result due to the small number of CL quasars thus far discovered. Under the simple assumption that transitions of either variety are equally likely both intrinsically and observationally (i.e. neglecting selection effects), the probability of observing three quasar-like to galaxy-like transition and no cases of the reverse is approximately 0.5$^3$ = 12.5\% from the binomial distribution. In the absence of selection effects, it thus appears unlikely for our search to randomly yield only quasar-like to galaxy-like CL quasars if the reverse transition is equally likely, though the statistics are not yet compelling. Recent searches for CL quasars using alternative approaches based on photometric light curves have now uncovered objects interpreted as galaxy-like to quasar-like CL quasars \citep{macleod15}.

\end{appendix}

\end{document}